\documentclass[prodmode, acmtoms]{acmsmall}

\usepackage[T1]{fontenc}
\usepackage[utf8]{inputenc}
\usepackage{csquotes}
\usepackage[mode=buildmissing]{standalone}

\usepackage{amsmath}
\usepackage{amssymb}
\usepackage{ebproof}
\usepackage{mathtools}
\usepackage{bm}
\usepackage{moresize}
\usepackage[nounderscore]{syntax}

\usepackage[ruled]{algorithm2e}
\usepackage[listings,most]{tcolorbox}
\usepackage{etoolbox}

\SetAlFnt{\small}
\SetAlCapFnt{\small}
\SetAlCapNameFnt{\small}
\SetAlCapHSkip{0pt}
\IncMargin{-\parindent}

\usepackage{booktabs}
\usepackage{tikz}
  \usetikzlibrary{calc, backgrounds, shapes, positioning}
\usepackage{graphicx}

\usepackage{listings,lstautogobble,xcolor,beramono}

\definecolor{diffstart}{HTML}{808080} % Gray
\definecolor{diffincl}{HTML}{008000}  % Green
\definecolor{diffrem}{HTML}{FF4500}   % OrangeRed

\lstdefinestyle{default}{
  tabsize=2,
  numbers=left,
  breaklines=true,
  postbreak=\raisebox{0ex}[0ex][0ex]{\ensuremath{\color{darkgray}\hookrightarrow\space}},
}
\lstdefinelanguage{PPML}{
  language=Fortran,
  morekeywords={%
    fields, particles, initialize, solve, integrate, on, from, to, by, exchange ghosts,%
    print, end, distribution, displaced, foreach, with, position, if, import, external, then, client, rhs%
  },
  morecomment=[l]{!},
  moredelim=[is][keywordstyle]{||}{||}
}

\lstdefinelanguage{PPME}{
  language=Java,
  keywordstyle=\color{blue!50!black}\bfseries,
  morekeywords={foreach, particle, ode, method, on, end},
  stringstyle=\color{green!50!black}\bfseries
}

\lstdefinelanguage{diff}{
  basicstyle=\ttfamily\small,
  morecomment=[f][\color{diffstart}]{@@},
  morecomment=[f][\color{diffincl}]{+\ },
  morecomment=[f][\color{diffincl}]{>\ },
  morecomment=[f][\color{diffrem}]{-\ },
  morecomment=[f][\color{diffrem}]{<\ },
}

\def\inline{\lstinline[basicstyle=\ttfamily\small,mathescape=true,breaklines=false]}
\newcommand{\arrowOp}[2]{\ensuremath{#1{\to}#2}}

\newcommand*\circled[1]{\tikz[baseline=(char.base)]{
            \node[shape=circle,draw,inner sep=0.5pt] (char) {#1};}}

\newcommand{\vect}[1]{\ensuremath{\mathbb{V}\langle #1 \rangle}}

\newcommand{\prop}[1]{\ensuremath{\mathcal{E}\langle #1 \rangle}}
\newcommand{\integer}{\ensuremath{\mathbb{Z}}}
\newcommand{\bool}{\ensuremath{\mathbb{B}}}
\newcommand{\real}{\ensuremath{\mathbb{R}}}

\newcommand{\improvementsFigure}[3]%
  {%
    \centering\footnotesize
    \begin{tabular}{lr}
      \toprule
      \textbf{Input Error} & \textbf{Output Error} \\
      \midrule
      $#1$ & $#2$ \\
      \midrule
      \multicolumn{2}{c}{
        \begin{minipage}{.85\textwidth}
          \footnotesize
          #3
        \end{minipage}
      } \\
      \bottomrule
    \end{tabular}
  }

\newcommand\NoBibDot[1]{}

\makeatletter
\newenvironment{btHighlight}[1][]
{\begingroup\tikzset{bt@Highlight@par/.style={#1}}\begin{lrbox}{\@tempboxa}}
{\end{lrbox}\bt@HL@box[bt@Highlight@par]{\@tempboxa}\endgroup}

\newcommand\btHL[1][]{%
  \begin{btHighlight}[#1]\bgroup\aftergroup\bt@HL@endenv%
}
\def\bt@HL@endenv{%
  \end{btHighlight}%
  \egroup
}
\newcommand{\bt@HL@box}[2][]{%
  \tikz[#1]{%
    \pgfpathrectangle{\pgfpoint{1pt}{0pt}}{\pgfpoint{\wd #2}{\ht #2}}%
    \pgfusepath{use as bounding box}%
    \node[anchor=base west, fill=orange!30,outer sep=0pt,inner xsep=1pt, inner ysep=0pt, rounded corners=3pt, minimum height=\ht\strutbox+1pt,#1]{\raisebox{1pt}{\strut}\strut\usebox{#2}};
  }%
}
\makeatother

\newboolean{showcomments}
\setboolean{showcomments}{false}
\ifthenelse{\boolean{showcomments}}
{ \newcommand{\mynote}[3]{
   \fbox{\bfseries\sffamily\scriptsize#1}
   {\small$\blacktriangleright$\textsf{\emph{\color{#3}{#2}}}$\blacktriangleleft$}}}
{ \newcommand{\mynote}[3]{}}
\newcommand{\tn}[1]{\mynote{Tobias}{#1}{red}}

\newcommand{\revii}[1]{#1}

\acmVolume{V}
\acmNumber{N}
\acmArticle{A}
\acmYear{YYYY}
\acmMonth{11}
\setcopyright{acmlicensed}

\begin{document}

% --------------------------------------------------------------------------- %
% Preface (abstract, classification, author details, etc.)
% --------------------------------------------------------------------------- %

% --------------------------------------------------------------------------- %
% Title, Author, and Page Heads
% --------------------------------------------------------------------------- %
\title{A Domain-Specific Language and Editor for Parallel Particle Methods }

\author{SVEN KAROL$^1$, TOBIAS NETT$^1$, JERONIMO CASTRILLON$^1$ and IVO F.
SBALZARINI$^{1,2}$ 
\affil{$^1$: Technische Universit\"at Dresden, Faculty of Computer Science, Dresden, Germany\\
$^2$: Center for Systems Biology Dresden, Max Planck Institute of Molecular Cell Biology 
and Genetics, Dresden, Germany}}

% --- End of Author Metadata ---

\markboth{S. Karol, T. Nett, J. Castrillon and I. F. Sbalzarini}{A Domain-Specific Language and Editor for Parallel Particle Methods}

%
% At a minimum you need to supply the author names, year and a title.
% IMPORTANT:
% Full first names whenever they are known, surname last, followed by a period.
% In the case of two authors, 'and' is placed between them.
% In the case of three or more authors, the serial comma is used, that is, all author names
% except the last one but including the penultimate author's name are followed by a comma,
% and then 'and' is placed before the final author's name.
% If only first and middle initials are known, then each initial
% is followed by a period and they are separated by a space.
% The remaining information (journal title, volume, article number, date, etc.) is 'auto-generated'.
\acmformat{Sven Karol, Tobias Nett, Jeronimo Castrillon and Ivo F.~Sbalzarini. 2017. A Domain-specific Language and Editor for Parallel Particle Methods}

% --------------------------------------------------------------------------- %
% Abstract
% --------------------------------------------------------------------------- %

\begin{abstract}
% JCM: abstract on 200 chars
Domain-specific languages (DSLs) are of increasing importance in scientific high-performance computing to reduce development costs, raise the level of abstraction and, thus, ease scientific programming. However, designing DSLs is not easy, as it requires knowledge of the application domain and experience in language engineering and compilers. Consequently, many DSLs follow a weak approach using macros or text generators, which lack many of the features that make a DSL comfortable for programmers. Some of these features---e.g., syntax highlighting, type inference, error reporting---are easily provided by language workbenches, which combine language engineering techniques and tools in a common ecosystem. In this paper, we present the Parallel Particle-Mesh Environment (PPME), a DSL and development environment for numerical simulations based on particle methods and hybrid particle-mesh methods. PPME uses the Meta Programming System (MPS), a projectional language workbench. PPME is the successor of the Parallel Particle-Mesh Language, a Fortran-based DSL that uses conventional implementation strategies. We analyze and compare both languages and demonstrate how the programmer's experience is improved using static analyses and projectional editing\revii{, i.e., code-structure editing, constrained by syntax, as opposed to free-text editing}. We present an explicit domain model for particle abstractions and the first formal type system for particle methods.
\end{abstract}

% --------------------------------------------------------------------------- %
% ACM Classification
% --------------------------------------------------------------------------- %
%
% ACM publications are classified according to the ACM Computing Classification Scheme (CCS). CCS codes are used both in the typeset version of the publications and in the metadata in the various databases. Therefore you need to provide both TEX commands and XML metadata with the paper.


%
% The command \terms is obsolete; we no longer use “General Terms” line.
%
%\terms{Design, Languages}

% --------------------------------------------------------------------------- %
% Keywords
% --------------------------------------------------------------------------- %
%
% The “Additional Keywords and Phrases” item on the title page is provided by the \keywords declaration, listed alphabetically.
% There is no prescribed list of “additional keywords;” use any that you want.

\begin{bottomstuff}
\textit{Preprint}. This work is partly supported by the German Research Foundation (DFG) within the Cluster of Excellence “Center for Advancing Electronics Dresden” (EXC 1056). 
\end{bottomstuff}
% --------------------------------------------------------------------------- %
% Generate title
% --------------------------------------------------------------------------- %
\maketitle

\section{Introduction} The emergence of massively parallel hardware architectures, 
such as different kinds of multi- and many-cores, general-purpose GPUs, and FPGAs in
scientific high-performance computing (HPC) has led to the development of new (or the
renovation of old)  programming models, paradigms, languages, and standards.
Standardized interfaces such as the Message Passing Interface
(MPI)~\cite{MPI30}, OpenMP~\cite{openmp_standard},
OpenACC~\cite{openacc_standard}, or hardware-specific low-level programming languages
such as CUDA~\cite{cuda_standard} made their way into HPC programming as libraries,
language extensions, or compilers. However, using these tools efficiently in
scientific programming requires in-depth knowledge of the underlying HPC
architecture, the development of parallel applications, and numerical simulation
methods. Hence, the achievable level of abstraction remains rather low, which is a
well-known problem in scientific
programming~\cite{hannay_how_2009,wilson_wheres_2006}, causing the ``knowledge gap''
in program efficiency~\cite{Sbalzarini:2010}.

To address this gap, scientific libraries and domain-specific languages (DSLs)
have evolved into an important tool set in scientific HPC. However, most of the scientific DSLs are built on
rather conventional technology such as macros, templates, and/or parser generators. 
In recent years, more sophisticated tools have been proposed,
frequently referred to as \emph{language
workbenches}~\cite{fowler_language_workbenches_2005,erdweg_workbenches_2013}, which
enable developers to more easily create and integrate DSLs following a model-centric
approach. A main driver behind the rising interest in such tools is the paradigm of
\emph{language-oriented programming}~\cite{Ward1994}, where DSLs are created to
describe and solve software problems instead of using general-purpose languages, with
the goal of increased productivity and better maintainability through abstraction. 
Models are the central paradigm that is edited by users and automatically
transformed or interpreted by the workbench tooling. From this integrative idea,
major advantages over conventional approaches arise. Most language workbenches
provide configurable features known from professional programming environments, such
as automatic code completion, refactoring, and syntax highlighting.
Moreover, they typically provide a collection of internal, tailor-made specification
languages that address common concerns in language development, e.g., languages for
pretty-printing, rewriting, parsing, and code analysis or generation.

These tools were not used when designing the Parallel Particle Mesh library (PPM) and the Parallel Particle Mesh Language
(PPML) as a library and a DSL for large-scale scientific HPC using particle-mesh
abstractions~\cite{Sbalzarini2006a,Sbalzarini:2010,Awile:2013,Awile:2013a}. Instead, PPML was
implemented conventionally as an \emph{internal DSL}, embedded into Fortran 2003.
However, as PPML does not have a completely integrated language model, it is
\revii{difficult} to maintain, debug, extend, or optimize PPML
programs~\cite{Karol2015a}. To improve on these issues, we developed the PPM
Environment (PPME) as a\revii{n} Integrated Development Environment (IDE) for particle-mesh methods.
Based on the Meta Programming System (MPS)~\cite{Dmitriev2004,MPS32UsersGuide}, a language workbench that closely
follows the ideas of language-oriented programming, \revii{PPME provides an additional layer of abstraction on top of 
the PPML stack (cf. Figure~\ref{fig:ppme-access_layer} in Section~\ref{sec:implementation_use_cases})}. 
\revii{In contrast to text-based language workbenches, MPS relies on \emph{projectional editing} 
where users directly operate on a rendered, form-like ``projection'' of the program~\cite{feiler_incremental_1981}. This enables advanced rendering of tables and mathematical equations inlined with normal program code.}
\revii{Due to its underlying principles, we believe that MPS is an excellent 
platform to design languages that address the ``knowledge gap'' and raise the level of abstraction in scientific programming.}

In this paper, we present PPME as the first IDE for high-performance particle
simulations.  We introduce a complete language model that provides the corresponding
abstractions and paves the way for further domain-specific analyses. As an
important first analysis, we implement a static type-inference engine, supported
by a formal type system. Furthermore, we demonstrate the advantages of this 
approach and of using language workbenches for numerical optimizations by integrating a
mechanism for error-reduction in floating-point expressions. 

The remainder of this paper is structured as follows: Section~\ref{sec:particle_methods} briefly introduces the background of particle
methods. 
%Afterwards, in \jc{I'm not a friend of this time adverbs}
% SK: I like them, but feel free to remove them where you like.
Section~\ref{sec:ppml} provides a more detailed overview and analysis of the current
PPML implementation and tool flow. The language model and type system of PPME are 
discussed in Section~\ref{sec:model-and-type-system}. Section~\ref{sec:implementation_use_cases}
gives an overview of PPME's coarse-grained architecture and implementation, and 
presents three case studies. In Section~\ref{sec:analysis}, we show how optimizations
can be added by integrating an external analysis tool using domain-specific information.
A qualitative evaluation of our work is given in Section~\ref{sec:evaluation}.
Finally, we discuss related work in Section~\ref{sec:related_work}, and Section~\ref{sec:conclusions} 
concludes the paper.

\section{Particle Methods}
\label{sec:particle_methods}
Particle methods provide a universal approach for numerical simulations in scientific computing. 
In contrast to other simulation frameworks, such as finite element methods (FEM) or Monte-Carlo methods,
particle methods can simulate models of all four kinds: discrete/deterministic, discrete/stochastic, 
continuous/deterministic, continuous/stochastic \cite{Sbalzarini:2013}. In case of continuous models, particles correspond to 
discretization points. When discrete models are simulated, entities in a model are directly represented
by particles. In deterministic simulations, particle positions and properties evolve according to deterministic
interactions between particles, whereas in stochastic models, these interactions are probabilistic.

In general, particles are zero-dimensional point-like objects characterized by a collection of properties 
of arbitrary types and a position in any space given as a vector whose length corresponds to the dimension
of that space. While a particle always has a position, its list of properties may grow or shrink in the course of 
a simulation. 
As an example of a discrete particle, consider a car on a street. The car's position may correspond to its 
GPS coordinates on a map while its properties could be velocity, the driver's age, number of passengers,  
or the color of the car. Other examples may be a pixel of an image (i.e., in a discrete space) or a discretization point of a continuous mathematical field, 
where the space is continuous.

Particles can \emph{interact} pairwise with other particles and they can \emph{evolve}. Evolving means
that a particle's position and/or properties change due to its own state and/or the states of other particles in the domain.
The influence of other particles is due to the interactions, which may yield a contribution to the change.
Hence, in pseudo code, the essential ingredients of particle objects can be described as shown in 
Figure~\ref{list:particles_structure}.%
{\addtolength{\abovecaptionskip}{-4mm}%
%\addtolength{\belowcaptionskip}{-1pt}%
\begin{figure}[th]%
\begin{lstlisting}[basicstyle=\footnotesize\ttfamily,numbers=none,
keywords={class,method,struct,vector},belowskip=0pt,aboveskip=0pt,
framesep=0pt,rulesep=0pt,frame=none,mathescape=true]
class PARTICLE {
    vector(space-dimension) :: position $\vec{x}$, positionChange $\Delta \vec{x}$
    struct :: properties $\vec{\omega}$, propertiesChange $\Delta \vec{\omega}$
    
    method [vector $\vec{K}_x$, struct $\vec{K}_\omega$]  interact(PARTICLE $q$)
    method evolve()
}
\end{lstlisting}%
\caption{Particle declaration and properties in pseudo code.}%
\label{list:particles_structure}%
\end{figure}%
}
Using this very basic interface, the evolution of a particle may depend on the position vector $\vec{x}$
and the list of properties  $\vec{\omega}$ of the particle itself, as well as the values of all other particles
in the system. If we assume that all particles influence each other, in the most general form, the 
 changes for any particle $p$ in the system can be described abstractly by Eq.~\ref{eq:pmgeneric}:
\begin{equation}
\begin{bmatrix} \Delta \vec{x}_p \\ \Delta \vec{\omega}_p \end{bmatrix} = \sum_{q=1}^{N} \begin{bmatrix} \vec{K}_x \\ \vec{K}_\omega \end{bmatrix} = \sum_{q=1}^{N} \vec{K}(\vec{x}_p, \vec{x}_q, \vec{\omega}_p, \vec{\omega}_q)\, .  \label{eq:pmgeneric}
\end{equation}
Here, $N$ is the total number of particles in the system, which may change over time (e.g., depending on a
boundary condition). $\vec{K}$ represents the \emph{interaction kernel} that encapsulates the 
computational model and corresponds to a mathematical representation of the \texttt{interact} method. 
\revii{Applied on position vectors $\vec{x}_p, \vec{x}_q$ and properties $\vec{\omega}_p, \vec{\omega}_q$}, the kernel produces the elementary changes $\vec{K}_x$ and $\vec{K}_\omega$ of the pairwise interactions 
between particles $p$ and $q$. The cumulative change for particle $p$ due to all interactions with other
particles is represented by two deltas, $\Delta \vec{x}_p$ and $\Delta \vec{\omega}_p $, which are used
by \texttt{evolve} to update the property values and position of $p$. 

In numerical simulations, updates of particle properties and positions occur at each time step. Thus, it is important
to evaluate the pairwise interactions efficiently. In the worst case, each particle interacts with each other 
particle, which leads to quadratic time complexity. However, typically, a particle only needs to interact with its
``neighbors'' within a finite range. In such cases, optimized data structures such as cell lists~\cite{Hockney_1988} 
exist, which allow for computing particle interactions in linear time (average complexity if particles are 
uniformly distributed). Nevertheless, the worst-case complexity remains quadratic if all particles are located
within the interaction range or the interaction range is the size of the domain. In these cases, efficient approximation algorithms
are available, e.g., the Barnes-Hut algorithm~\cite{Barnes_1986} and Fast Multipole Methods~\cite{Greengard_1987}.
Another way to address this problem is to use a hybrid particle-mesh approach, where interactions
of finite range are computed using particles, whereas interactions of infinite range are evaluated  
using mesh-based approaches~\cite{Hockney_1988}. 

% JC: breaking paragraph - Only about the kernel and discretizations.
The range of the particle--particle interactions is defined by the support of the interaction kernel $\vec{K}$. 
This kernel is the mathematical representation of the system to be simulated and encapsulates all application-specific details. 
When simulating discrete models, $\vec{K}$ corresponds to the pairwise interaction potential between the entities in the model, e.g., the inter-atomic force fields in a molecular-dynamics simulation. When simulating continuous models, such as partial differential equations (PDEs), $\vec{K}$ contains the discretized continuous or differential operators. In this case, the particles as discretization/colocation points at which the value of the continuous function is sampled. 

% On discretizations
Particle discretizations of differential operators in PDEs (i.e., the kernel $\vec{K}$) can be determined using a variety of classical approaches from numerical analysis ~\cite{Lucy_1977,liu1995reproducing,belytschko_1994,lancaster1981surfaces,broomhead1988radial,Degond_1989,Eldredge_2002} that are generic to arbitrary linear differential operators. They all have in common that the kernel $\vec{K}$ is pre-computed, usually analytically by hand, and then implemented in the discrete form. To free the scientific programmer from this analytical calculation, we here implement \revii{a method known as \emph{Discretization-Corrected Particle Strength Exchange (DC-PSE)}}~\cite{Schrader_2010}. 
DC-PSE is a general particle discretization framework where the discrete kernels are automatically computed at runtime. In addition, DC-PSE also shows superior stability and accuracy properties compared to other mesh-free discretization methods \cite{Schrader_2010,Reboux:2012,Schrader:2012,Bourantas:2016}.

% JCM: Not a friend of this sentence. If at all, then put it at the beginning of next section. 
%The next section introduces a classic reaction-diffusion problem and shows how it can be solved using 
%particle methods in PPM(L).
% SK: Why not, actually?
% JC: overdue answer ;-) ... I like connecting sentences, but rather at the beginning of (large) sections.
%    - With connections all over the place (e.g., beginning and end), the paper gets a tutorial-like feeling
%    - Sometime people jump to the start of a section, there the connection helps. At the end of the section, not so much. 

%%Moved into standalone because of strange
\begin{figure}[t]
\includegraphics{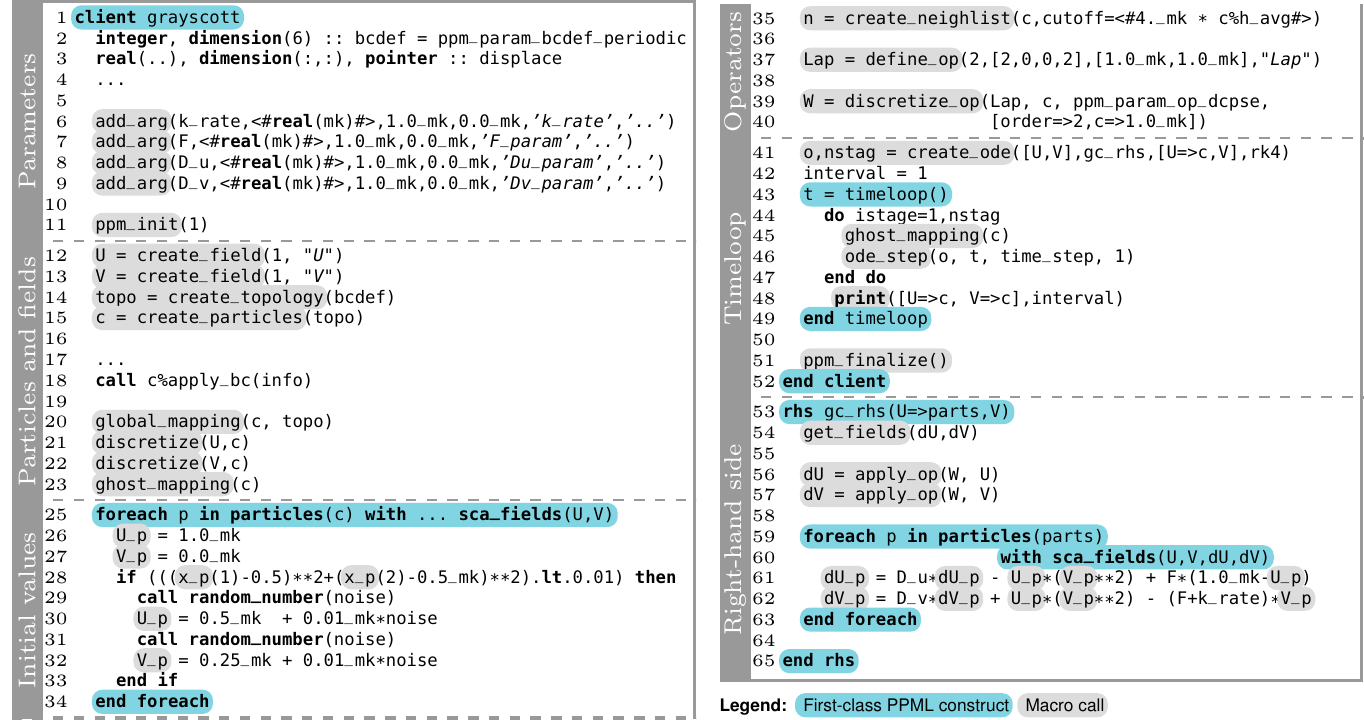}%
\label{fig:grayscott-ppml}%
\caption{PPML program to numerically solve the 2D Gray-Scott reaction-diffusion system on distributed-memory computer systems.}%
\end{figure}

\section{The Parallel Particle Mesh Language PPML}
\label{sec:ppml}
The Parallel Particle-Mesh Language (PPML)~\cite{Awile:2013,Awile:2013a} provides domain-specific abstractions to ease the development of distributed-memory 
particle-mesh simulations with the PPM HPC library~\cite{Sbalzarini2006a,Sbalzarini:2010,Awile:2010a}. The language is smoothly embedded into Fortran2003---the
implementation language of PPM. The major advantage of relying on PPML over using \revii{the PPM library} directly is that complex library protocols and parallelization code are hidden from the user. Instead, PPML provides first-class concepts for particle programming such as \emph{particles},
\emph{neighbor lists} (with optimized implementations in PPM), particle properties like \emph{vector} and \emph{scalar fields}, and \emph{differential-operator
 definitions}. Furthermore, particle-specific foreach loops are supported, as well as loops over discrete time steps.
 For high-performance parallelization, PPML supports distributed memory with message passing based on MPI. Several macro commands
 are provided to help handle the MPI setup, create topologies (i.e., decomposing the domain and assign subdomains to processes), distribute particles
 over these topologies, and exchange data at subdomain boundaries (cf.~\cite{Sbalzarini2006a}).

\subsection{A Simple Application Example}
\label{sub:a_simple_application_example}
To illustrate how parallel simulations can be implemented in PPML, we discuss an example of a Gray-Scott reaction-diffusion system, taken from the
PPML paper \cite{Awile:2013}.
A Gray-Scott reaction-diffusion system describes \revii{the concentrations (in normalized dimensionless units)} of two chemicals $u$ and $v$ that react with each other
and diffuse~\cite{Gray:1983}. The process can be described by a system of two partial differential equations that define the evolution of \revii{the chemicals' concentrations}, $u$ and $v$, over time:
\begin{align}
	\label{eq:u}\frac{\partial{u}}{\partial{t}}& = {D_u}\Delta{u} - uv^2 + F(1 - u)\\
	\label{eq:v}\frac{\partial{v}}{\partial{t}}& = {D_v}\Delta{v} + uv^2 - (F + k)v \, . 
\end{align}
Equation~\ref{eq:u} describes the time derivative of $u$ as a sum of three terms: First is the \emph{diffusion term} ${D_u}\Delta{u}$, where $D_u$ is a predefined diffusion constant
and $\Delta{u}$ the Laplacian (divergence of the gradient) of $u$. Second is the \emph{reaction term} $- uv^2$ defining how much of $u$ is converted to $v$ by the reaction. The last term in Eq.~\ref{eq:u} is the \emph{replenishment term}, defining how much of fresh $u$ is added to keep the reaction alive,
depending on a constant \emph{feed rate} $F$.
\revii{Equation~\ref{eq:v}} describes the time derivative of $v$ also as a sum of reaction, diffusion and, instead of a replenishment term, a \emph{diminishment term}. $D_v$ is the constant diffusion rate of $v$ and $- (F + k)v$ defines how much of $v$ is taken out (consumed) from the system, depending on
$F$ and a \emph{removal rate} $k$.%\jc{Isn't this too much details?}

This continuous model can be solved numerically using particle methods and PPML. To do so, $u$ and $v$ are discretized as particle properties and particles
are distributed over the entire domain. Furthermore, the differential operators (i.e., the Laplacians) need to be discretized according to the used method. After
providing initial values of $u$ and $v$ at particles, an approximate solution can be computed for a series of time steps.

%\sk{Should we call it solver, simulation, simulator or ...?}
Figure~\ref{fig:grayscott-ppml} shows the corresponding PPML \emph{client} program as a multi-part listing that highlights the different ingredients of the program. The first part on the left-hand side
of the figure contains variable
and constant declarations, the boundary condition, as well as declarations of external arguments  allowing users to parametrize the simulation. The second part declares $u$
and $v$ as scalar fields \inline{U} and \inline{V} and discretizes them over particles. Furthermore, a topology is created to distribute the particles on a computer cluster.
In the third part, the initial values of \inline{U} and \inline{V} are set using a PPML \inline{foreach} loop that iterates over all particles in the domain, by default assigning \inline{U} a value
of $1$ and \inline{V} a value of $0$. However, within a radius of $\sqrt{0.1}$ around the center, a small random amount of $v$ is added to start the reaction.
The fourth part (on top of the right-hand side of Figure~\ref{fig:grayscott-ppml}) contains the definition and discretization of the Laplace operator and initializes a particle neighbor list
with a specific \inline{cutoff}. \revii{The cutoff (i.e., the range of the particle--particle interactions) is set such that each particle interacts with all neighboring particles that are closer than four times the} average inter-particle distance. The remaining two portions of the figure specify the \inline{timeloop}, which
sequentially loops over the specified range of time steps, in each step evolving the solution by calling the PPM solver with the specified right-hand side, updating the
particle properties, exchanging data at inter-process boundaries, and printing the intermediate results to the file system. The right-hand side specification contains the
reaction-diffusion equations to be solved with an explicit invocation of the discretized Laplacian over both fields yielding respective vectors of intermediate results. Another PPML particle
\inline{foreach} loop computes the contributions of each individual particle using \LaTeX-like formula expressions, where underscores access individual particles.%\jc{Can you put numbers for the ``parts'' in the fig? and probably use code lines in the text?}
{\begin{figure}
\centering
\begin{minipage}[c]{0.24\textwidth}%
\centering\includegraphics[scale=0.6]{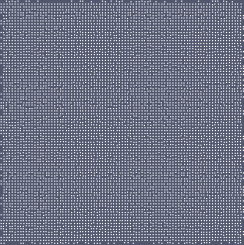}
\\\scriptsize $t=\emptyset$
\end{minipage}%
~
\begin{minipage}[c]{0.24\textwidth}%
\centering\includegraphics[scale=0.6]{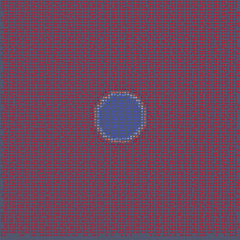}
\\\scriptsize $t=0$
\end{minipage}%
~
\begin{minipage}[c]{0.24\textwidth}%
\centering\includegraphics[scale=0.6]{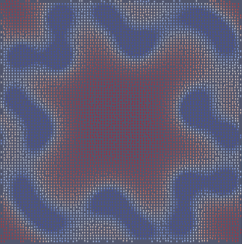}
\\\scriptsize $t_i>>t_0$
\end{minipage}%
~
\begin{minipage}[c]{0.24\textwidth}%
\centering\includegraphics[scale=0.6]{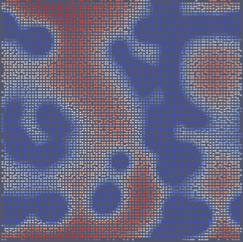}
\\\scriptsize $t_j>>t_i$
\end{minipage}%
\caption{Some intermediate results produced by the PPML Gray-Scott program.}%
\label{fig:grayscott-ppml-sim}%
\end{figure}}
Figure~\revii{\ref{fig:grayscott-ppml-sim}} shows some intermediate results over a small domain at different time steps that have been produced by this PPML program, choosing $k=0.051$, $F=0.015$, $D_u = 2\cdot 10^{-5}$, and $D_v = 10^{-5}$. For these parameters, the Gray-Scott system forms spatial patterns that are hypothesized since Alan Turing to be the chemical basis of biological growth and morphogenesis \cite{Turing:1952}.

\subsection{Advantages over Conventional Programming}
The program shown in Figure~\ref{fig:grayscott-ppml} nicely demonstrates some of the major benefits of DSLs in scientific HPC. Most of the boilerplate code
for instantiating PPM and managing parallelism with MPI is hidden from the developer. It is automatically generated by the PPML source-to-source compiler, emitting a plain Fortran
program, which is then compiled and linked with the PPM library by a standard Fortran compiler. The size ratio of the PPML source and the generated Fortran program
is 85:668, which means that the developer is freed from the burden of writing an extra 583 lines of boilerplate code.

The improved program readability is a further advantage of PPML over writing a plain Fortran program. Thanks to built-in domain-specific concepts and other specialized constructs, such as
particle loops and underscore accessors, the program is more declarative and thus \revii{more} readable, so that other domain experts can easily understand \revii{it}.
Finally, PPML was designed as an extensible \revii{language, apparently embedded into Fortran as a host language}. This way, it circumvents one of the major 
obstacles of using closed, stand-alone DSLs, namely a lack of expressiveness that may prevent facets of a problem to be described using the abstractions at hand. If a problem cannot be described properly using PPML, developers \revii{can always use} plain Fortran (e.g., lines 28, 29, and 31) or they may define additional PPML macros.
\revii{However, since PPML lacks a well-defined interface between the DSL and its host language, it is difficult to properly analyze the code and derive  context information from it. This lack of context largely prevents automatic compile-time code optimization in PPML. 
Such optimizations are easier with non-embedded DSLs, where language interfaces need not to be considered.}
%JC: Here we say that embedding is good in case of lack of expressiveness. What is then our argument for a non-
%embedded one? Are we sure there we do not leave unsupported ``facets''?
%SK: Yes, embedding is good if it is done in the right way and does not destroy the benefits of the DSL.

\subsection{Limitations in the Current PPML Design}
The current design and implementation of PPML has some limitations that hamper code optimization and debugging. The most important limitation is that the language is
not based on a formal domain model, which would enable reasoning about PPML programs to automatically check consistency, e.g., using a formal type system. Moreover,
the language syntax is underspecified. Similar to an island grammar~\cite{moonen_generating_2001}, only some parts of the language are modeled explicitly,  while others remain undefined.
Consider Figure~\ref{fig:grayscott-ppml} again. Parts of the program that are recognized by the PPML \revii{source-to-source} compiler are highlighted in either gray (macro calls including list of arguments)
or blue (first-class language constructs). These are the structural ``islands'' in the sense of an island grammar, while the non-highlighted parts are ``water'', i.e., parts of the program that are treated as a list of
characters \revii{that do not provide any additional information to the PPML compiler}.  
This \revii{fragmentary view on the code} allows for only shallow analyses of input programs \revii{during the preprocessing phase}, leaving most syntax and type errors undetected \revii{so that these are inherited by the generated program}. 
\revii{If such a program is then fed into a Fortran compiler to produce an executable,
the compiler will detect these issues and associate them with the preprocessed code. However, the developer has neither seen nor written this automatically generated Fortran code and can therefore not trace back the errors to his PPML program. Debugging PPML programs is therefore unpractical.} 
Even worse, some problems manifest only during or after execution by causing unintended
results, e.g., through an unsuitable argument or wrong arrangement of calls.

By leveraging domain knowledge, one can define
a complete domain model for the language implementation \revii{so that syntactic elements and semantic relations can be identified and used in a compiler.  
This way, syntactic problems and problems related to the language semantics can be detected early when processing DSL code (e.g., differential operators that are not used in an equation or have the wrong type of operand).}
This further provides the \revii{groundwork} for improving user experience by adding features known from integrated development environments (IDEs) (e.g., syntax highlighting and code completion) and from
compilers (e.g., optimizations such as tiling, program variants, or expression rewrites that only become possible through the additional information). \revii{This is enabled here by formulating a domain metamodel and a formal type system for the application domain of particle methods.}
%
% JCM: Again: a bit against these kinds of sentences. If I jump to the next section as a reader, because that's the one I want to read, I wouldn't get to read this sentence ;-)
%In the following sections, we show how we achieved these goals in PPME, the successor of PPML, using language-workbench technology.
% SK: I think this is a matter of style. I normally write these sentences to somehow glue sections together. But
% I also do not have a problem leaving them out if this is done consistently.

\section{A Domain Metamodel and Type System for Particle Methods}
\label{sec:model-and-type-system}
A crucial step to overcome the current shortcomings of PPML is to develop a metamodel that 
structurally represents the domain of particle methods.  This ``domain model'' enables compile-time reasoning by providing a
structural basis for developing PPM programs. In this section, we describe such a model for particle methods, 
% JCM: Sentence below soudns quite like a report ``that we have developed''. I changed it for ``here is what it is'' ;-)
%we first discuss excerpts of the metamodel that we have
%developed to represent particle simulations in PPME. 
resulting in a static type system.

\subsection{Domain Metamodel}
\label{sec:domain-metamodel}%
\begin{figure}%
\includegraphics[width=\textwidth]{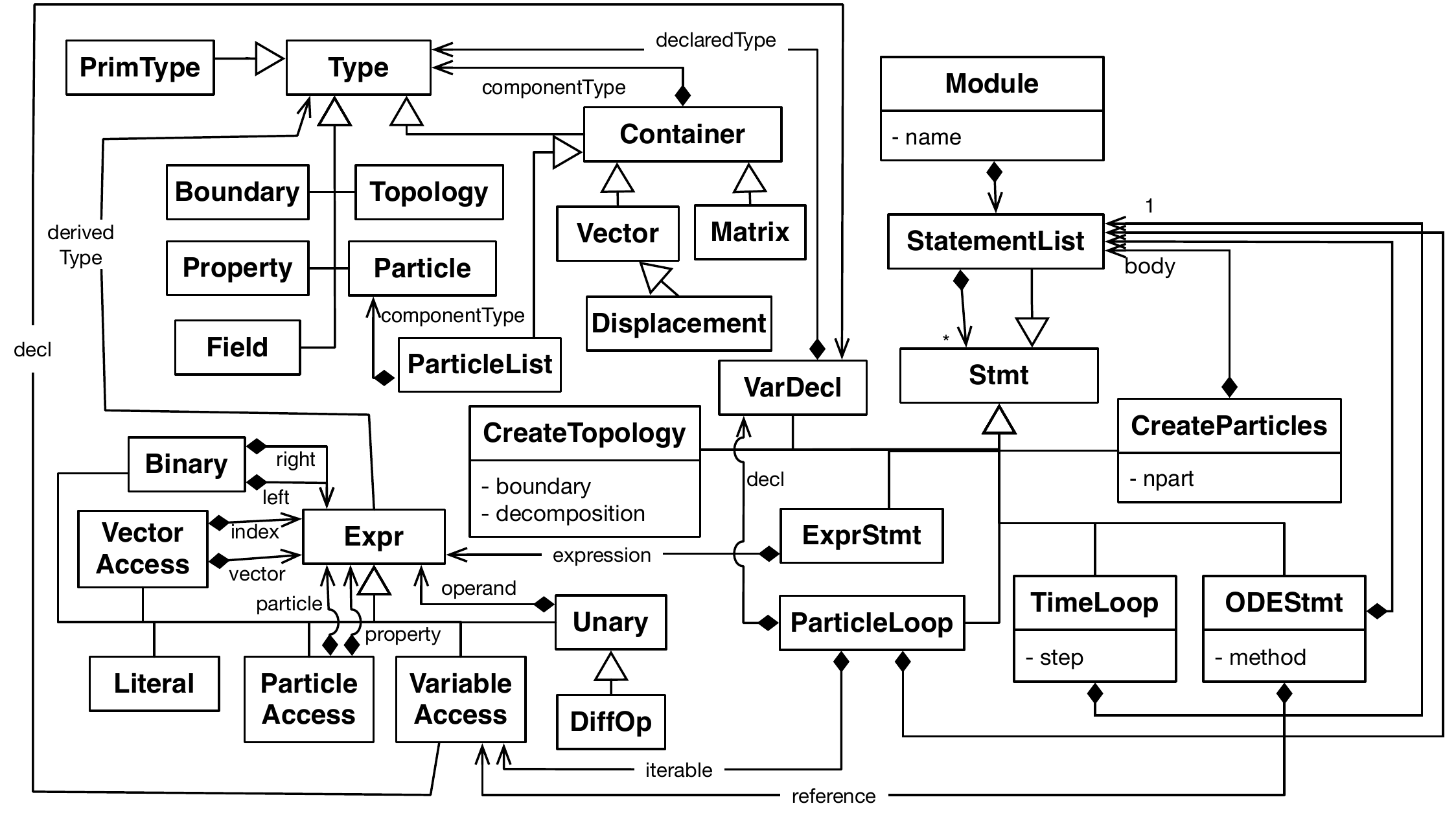}%
\caption{A metamodel to describe the domain of particle methods.}%
\label{fig:domain-model}%
\end{figure}%
The current PPML only supports a small fraction of the concepts that 
constitute a complete model. For instance, it provides constructs for defining computations over properties 
of particles in a domain, as well as loops for defining numerical simulations over a series of discrete time steps.
However, since these concepts are only specified partially in PPML, it is not possible to reason about the actual
computation steps. 

%\jc{Similar to tex-comment above: I would suggest suggest to change a bit the writing. Instead of ``our metamodel needs to consider'' or ``it should provide means'', write ``our metamodel considers'' and ``... provides means''. It otherwise sounds like a wish-list.}
\revii{In contrast to PPML, we propose a complete metamodel for particle 
methods that captures particles, particle data and computations over these
data. 
More precisely, it provides a means to specify particle data
structures and properties. 
It supports high-level statements that group transformations of particle data, which allow modifying data within a certain scope. 
Furthermore, the model includes mathematical expressions to describe the actual governing equations. 
Finally, our metamodel defines the set and structure of potential data types supported in PPME.}

% JC: Next as separate paragraph - (issue 13)
Figure~\ref{fig:domain-model} shows an excerpt of the model as a UML class diagram. We derived its major ingredients from our experience of developing HPC applications with PPML. A single top-level concept \inline{Module} contains a list of 
statements that describe a particle-based simulation. A statement, for instance, can be a composite \inline{Timeloop} 
or an elementary command for creating a topology. Single computation steps are described using expression statements
(\inline{ExprStmt}). Expressions are binary or unary arithmetic and logical expressions that access constants,
particle properties, and collections of particles. Furthermore, expressions can define differential operators (\inline{DiffOp}s)
evaluated over particle properties, e.g., the Laplacian. 
Our model also provides a simple \inline{Type} hierarchy with a set of primitive types (e.g., integers) and built-in domain-specific 
types. The domain-specific types can be atomic (e.g., \inline{Particle} or \inline{Topology}) or container types such as 
\inline{Vector} or \inline{ParticleList}. 

When considering the model in Figure~\ref{fig:domain-model}, it is evident that it only provides structural
information, since it is limited to \emph{potential} relations between objects and types, such as inheritance, 
composition, and reference. Hence, while an instance of this model (i.e., a concrete specification of a particle method) 
may be syntactically correct, it may not fulfill requirements that are not specified in the model. 
Properties that impose such additional constrains need to be formulated as supplementary rules that derive additional information 
or check consistency of a specification~\cite{buerger_reference_2011}. 
As an example of where additional information needs to be derived, consider the \inline{decl} reference that associates 
an access of a variable with a corresponding declaration. This is required because users would not ``draw'' the corresponding 
connection but just ``use'' the variable via its name. 
Another important analysis computes the types of expressions, as, for instance, represented by the \inline{derivedType} reference 
in the model, which associates an expression with a specific type. The rules that define this analysis can be captured 
conveniently using a formal type system~\cite{Plotkin1981}. 
% JCM: And again, but I leave it up to you ;-) 
%In the following, we present an according domain-specific 
%formal type system for our model, which was implemented in PPME.

% vim: set ts=2 sw=2 sts=2 tw=85:
% vim: set wrap breakindent:
% !TEX root = master.tex        --- atom

\subsection{Types and Dimensions} % (fold)
\label{sec:types_units}

  Based on the domain model, we present a static type-inference mechanism,
  which relies on a formal type system for particle abstractions. The error
  detection capabilities resulting from the hierarchy of types and inference rules
  are key to constructively improve code quality of simulations written
  in PPME, as it detects errors at compile time and provides meaningful feedback to
  the developer. In addition, we
  present an optional unit calculus extension to the type system. This can be used to perform automatic consistency checks of expressions.
% subsection types_units (end)

\subsubsection{Type Hierarchy} % (fold)
\label{sub:type_hierarchy}
  %The underlying language workbench allows to freely define custom types by language concepts.
  The type hierarchy is built around the metamodel shown in
  Figure~\ref{fig:domain-model}, i.e., all types derive from \inline{Type} as a
  common supertype.  The type system can be divided into two parts: a base type
  system and a domain-specific extension.

  \newcommand{\TP}[2][Xdummy]{\ensuremath{\mathit{#2}\ifstrequal{#1}{Xdummy}{}{\langle\mathit{#1}\rangle}}}

  % -- Base Types -----------------------------------------------------------
  The base type system consists of a set of primitive types
    $\mathcal{P} = \{ \TP{String}, \TP{Boolean}, \TP{Real}, \TP{Integer} \}$
  and type-inference rules over this set.
  %The set $\mathcal{P}$ of primitive types contains the well-known types \inline{String} for character sequences, \inline{Boolean} for truth values, \inline{Integer} for integers, and \inline{Real} as an %abstraction for the real numbers.
  % \[ \mathcal{P} = \{ String, Boolean, Real, Integer \} \]
  %Additionally, PPME knows two kinds of parameterized types for vectors and matrices, denoted by \lstinline{Vector<X>} and \lstinline{Matrix<X>} where \inline{X} is a \emph{type parameter}.
  Additionally, $\mathcal{C} = \{ \TP[X]{Vector}, \TP[X]{Matrix} \}$ represents a set
  of container types for matrices (i.e., tensors of rank 2) and vectors (i.e., tensors of rank 1) with components of type $X$.
   %A vector and matrix represent one-dimensional and two-dimensional arrays with components of type \inline{X}, respectively.
   %These polymorphic types are joined under an abstract \lstinline{ContainerType} in the implementation and are denoted by the set $\mathcal{C}$.
  %\[ \mathcal{C} = \{ Vector\langle{}X\rangle{}, Matrix\langle{}X\rangle{} \} \]
  The set of base types $\mathcal{T}_\mathit{Base} = \mathcal{P} \cup \mathcal{C}$ is
  composed of primitive types $\mathcal{P}$ and container types $\mathcal{C}$.

  % -- Domain-speicfic Types ------------------------------------------------
  % The domain-specific extensions to the type system are located in \lstinline{de.ppme.core}.
  These basic types are complemented by domain-specific types for particle methods,
  i.e., types that represent particles, particle lists, and different kinds of particle
  properties. These are:
  $\mathcal{D} = \{ \TP{Particle}, \TP{ParticleList}, \TP{Field}, \TP{Property},
  \TP{Displacement} \}$.
  Furthermore, the boundary of the simulation domain and the data-distribution topology of the underlying PPM framework are captured in the set $\mathcal{O}=\{ \TP{Topology}, \TP{Boundary} \}$. The set
  $\mathcal{T}_\mathit{PPM} = \mathcal{D} \cup \mathcal{O}$ of domain-specific types
  is then composed of $\mathcal{D}$ and $\mathcal{O}$.
  %
  %\[ \mathcal{T}_\mathit{PPM} = \mathcal{D} \cup \mathcal{O} = \{ \TP{Particle}, \TP{ParticleList}, \TP{Field}, \TP{Property}, \TP{Displacement} \} \cup \{ Topology, ... \} \]

  Finally, $\mathcal{T} = \mathcal{T}_{Base} \cup \mathcal{T}_{PPM}$ denotes the set
  of all types in PPME. Note that this way of constructing $\mathcal{T}$ indicates the
  flexibility of language implementations in modern language workbenches like MPS and
  language-oriented programming in general. This fundamental type hierarchy can be
  extended in the future, adding new domain-specific concepts.

  %SK: I think we do not need to discuss this further
  %This enables a
  %continuous refinement of the given implementation towards new use cases.\sk{Discuss
  %again in impl section.}\jc{That last point is kind of unclear to me.} \tn{e.g.,
  %\inline{Image} as special particle list, or \enquote{improving} existing types
  %(simply added size to vectors)}
% subsection type_hierarchy (end)

\subsubsection{Syntax of Expressions} % (fold)
\label{sub:syntax_of_expressions}

\revii{
  In PPME the standard set of expressions well-known by programmers is extended
  by domain-specific operations and expressions tailored for the domain.
  Figure~\ref{fig:expr-grammar} presents the syntax of expressions in PPME as
  production rules of a context-free grammar.
  Note that some domain-specific expressions (e.g., differential operators) are
  only available in a specific context.

  \tn{Differential operators as domain-specific expressions (e.g., Laplace and
  Jacobi) are restricted to the context of RHS statements. A right-hand side
  statement is not an expression, thus, it is not covvered by this section.}

  \tn{PPME-Lang itself is not context free, and I don't know how much sense it
  would make trying to represent the complete syntax of PPME in a grammar. The
  review comment addresses section 4.2.2 which only deals with expressions}
}

  \begin{figure}
    \centering
    \footnotesize
    \begin{minipage}[t]{.4\textwidth}
      \begin{grammar}
        <expr> ::= <expr> ( `&&' | `||' ) <expr>
        \alt <expr> ( `==' | `!=' ) <expr>
        \alt <expr> ( `<' | `>' | `<=' | `>=' ) <expr>
        \alt <expr> ( `+' | `-' ) <expr>
        \alt <expr> ( `*' | `/' | `^' ) <expr>
        \alt <unaryExpr>

        <unaryExpr> ::= `-' <unaryExpr>
        \alt `!' <unaryExpr>
        \alt \lit{$\sqrt{}$} <unaryExpr>
        \alt <primaryExpr>

        <varAccess> ::= Identifier
      \end{grammar}
    \end{minipage}
    \begin{minipage}[t]{.4\textwidth}
      \begin{grammar}
        <primaryExpr> ::= <literal>
        \alt `(' <expr> `)'
        \alt <varAccess>
        \alt <particleAccess>
        \alt <arrayAccess>

        <literal> ::= IntegerLiteral
        \alt RealLiteral
        \alt StringLiteral
        \alt BooleanLiteral

        <particleAccess> ::= <expr> \lit{$\rightarrow$} Identifier

        <arrayAccess> ::= <expr> `[' <expr> `]'
      \end{grammar}
    \end{minipage}
    \caption{\revii{Syntax of expressions in PPME.}}
    \label{fig:expr-grammar}
  \end{figure}

  \paragraph{Basic Syntactic Sets} % (fold)
  \label{par:basic_syntactic_sets}

  The basic syntactic sets in PPME are comprised mainly of \emph{literals} for
  primitive types and \emph{variables} (cf.
  Figure~\ref{tab:typesystem:basic_syntactic_sets}). Literals are typed in a natural
  way, e.g., integers have type $Integer$ and decimals have type $Real$. More
  complex sets can be derived from the basic syntactic sets for variables and
  literals. The abstract syntax of these derived syntactic sets is given by the form
  of expressions in PPME.
  \begin{figure}[tp]
    \centering
    \footnotesize
    \setlength{\tabcolsep}{0.5em}
    \begin{tabular}{lll}
      \toprule
      \textbf{booleans}  & $b$    & $b \in \mathbb{B} = \{ true, false \}$         \\
      \textbf{strings}   & $s$    & e.g., $s = \mathtt{"PPME"}$                  \\
      \textbf{integers}  & $n, m$ & $n, m \in \mathbb{N}$                          \\
      \textbf{reals}     & $r$    & e.g., $r = 3.14$ or $r = 6.62\mathrm{E}{-34}$\\
      \textbf{variables} & $v$    & $v \in Var = \{ a, b, \dots, x, x_2, \dots \}$
      \\
      \bottomrule
    \end{tabular}
    \caption{Basic syntactic sets and their notation.}
    \label{tab:typesystem:basic_syntactic_sets} % chkTex 24
  \end{figure}
  % paragraph basic_syntactic_sets (end)

  \paragraph{Unary Expressions ($\ominus(e)$, with $\ominus \in \{ -, !, \sqrt{} \}$)}
  \label{par:unary_expressions}

    PPME supports three unary operations, the unary minus $-e$, the logical not $!e$,
    and the square root $\sqrt{e}$. Obviously, this definition \revii{alone} allows for
    \enquote{nonesense} expressions such as taking the square root of a string. The
    remainder of this section therefore presents rules for well-formedness and type
    conclusion to prevent erroneous phrases.
    %   \[ \ominus \in \{ -, !, \sqrt{} \} \]
  % paragraph unary_expressions (end)

  \paragraph{Binary Expressions ($e_1 \otimes e_2$, where $\otimes \in \otimes_\mathit{arith} \cup \otimes_\mathit{logi} \cup \otimes_\mathit{rel}$)}
  \label{par:binary_expressions}

    Various binary operations are supported. First, PPME allows for typical arithmetic
    operators
      $\otimes_{arith} = \{ +, -, *, /, \text{\^{}} \}$
    Second, there are operators for the logical \emph{and} and \emph{or}
      ($\otimes_{logi} = \{ \&\&, || \} $).
    Third, the common relational operators are available
      ($\otimes_{rel} = \{ ==, !\!\!=, <, >, <=, >= \}$).
    % The union of these three categories yields the set of binary operations in PPME.
    As for unary operations, the type system will check well-formedness of binary
    expressions and decide on the resulting type.
    %\[ \otimes \in \otimes_{arith} \cup \otimes_{log} \cup \otimes_{comp} \]
    % \begin{align*}
    %   \otimes_{arith} \hfill&= \{ +, -, *, /, \text{\^{}} \}  \\
    %   \otimes_{log}   \hfill&= \{ \&\&, || \}                 \\
    %   \otimes_{rel}   \hfill&= \{ ==, !=, <, >, <=, >= \}
    % \end{align*}
  % paragraph binary_expressions (end)

  \paragraph{Domain-specific Operations}
  \label{par:domain_specific_operations}

    A strength of PPME is that domain-specific operations are seamlessly
    integrated into the language. They allow for concise notation of mathematical
    concepts, preserving the expressiveness of the mathematical notation. Following the
    domain model, fields and particle properties are defined on particle lists, and the language
    offers the syntactic concept \emph{particle list access} (PLA) to access these fields
    and properties. In a similar manner, the value of a field discretized over particles,
    or any other property of a specific particle, can be accessed via a \emph{particle access} (PA) operation.
    Given a particle list $ps$, a particle $p$ from this list, a field $f$ and a
    property $x$ both defined on $ps$, the access operations of field $f$ and property
    $x$ are represented by an arrow:%
    \[ \arrowOp{ps}{f}, \qquad \arrowOp{ps}{x}, \qquad \arrowOp{p}{f}, \qquad \arrowOp{p}{x} \]

    Intuitively, the result of a PLA is the whole field or property over the particle
    list. Additionally, PPME allows the developer to access the default properties of a
    particle, e.g., its position: $\arrowOp{p}{\mathit{pos}}$.
    Finally, there are notations for differential operators \revii{in the
    context of right-hand side statements}. Simulation developers can use these
    operators when simulating continuous models (e.g., PDEs), staying close to
    the mathematical notation. This includes, for example, the Laplacian
    ($\nabla^2 e$) used in the Gray-Scott reaction-diffusion example.
  % paragraph domain_specific_operations (end)

  \paragraph{Access Operations ($v[i]$, $m[i][j]$)} % (fold)
  \label{par:acces_operations}
    The language also offers means to access elements of array-like structures such
    as matrices and vectors. Access operations are denoted by square brackets
    containing the index to access. Similarly, elements of non-scalar particle properties
    can be accessed using the same notation. Let $ps$ be a particle list
    with a non-scalar field $f$, and $p \in ps$ a particle from $ps$, then
    $\arrowOp{p}{f}[i]$ denotes the access of the $i^{th}$ element of $f$ on particle $p$.
  % paragraph acces_operations (end)

% subsection syntax_of_expressions (end)

\subsubsection{Formal Type System}
\label{sec:type_rules}

  \revii{
    Every literal and variable in PPME has an associated type, and the type of derived expressions often depends on
    their arguments' types.
    The formal type system describes the conclusions that can be drawn from a PPME program over its types, by defining
    rules for well-formedness of typed expressions.
    For instance, the PLA operator can only be used on particle lists, which is ensured by the static type-inference
    mechanism.
    Overall, the type rules ensure that expressions behave as expected in the context of particle methods. 
  }

  % The description of the type system uses the following notational agreements and terminology.
  A \emph{typing environment} $\Gamma$ associates variable names $x$ and types $\tau$
  as a set of pairs $\langle x, \tau \rangle$, commonly written as $x : \tau$. A
  lookup of a variable's type is denoted by $\Gamma(x)$, where $\Gamma(x) = \tau$ if
  and only if the environment contains an entry for the variable $\langle x, \tau
  \rangle \in \Gamma$. Otherwise, $\Gamma(x)$ is undefined.
  We further define the subtype relationship: if $T$ and $S$ are types, then
  ${T}<{S}$ denotes that $T$ is a (direct) \emph{subtype} of $S$, and $S$ is a
  \emph{supertype} of $T$. ${T}{<^{*}}{S}$ is the reflexive transitive closure of $<$,
  that is, $S$ can be reached from $T$ in the type hierarchy. In the remainder, we
  use $\leq$ to refer to ${<^{*}}$.
  We follow the notation of~\cite{Clement1986}. That is, each type-inference rule
  defines the conclusion that can be drawn if all $n$ premises hold:
  \[
    \begin{prooftree}
      \Hypo{\mathit{premise}_1}
      \Hypo{\dots}
      \Hypo{\mathit{premise}_n}
      \Infer3{\mathit{conclusion}}
    \end{prooftree}
  \]
  As premises, we allow typings as well as other predicates, e.g., for specifying a
  subtype relation.
  %
  %-- type inference rules ------------------------------------------------
  \begin{figure}[t]
    %-- variables ---------------------------------------------------------
    %\begin{minipage}[t]{.50\textwidth}
    \footnotesize
    \begin{center}
      \textsc{Var}
      \begin{prooftree}%\label{eq:typerule_var}
        \Hypo{\Gamma(v) = \tau}
        \Infer1{\Gamma \vdash v : \tau}
      \end{prooftree}
      \textsc{VarDecl}
      \begin{prooftree}%\label{eq:typerule_vardecl}
        \Hypo{\phantom{G}}
        \Infer1{\Gamma \vdash \tau \text{ } x : \Gamma \cup \{x = \tau \}}
      \end{prooftree}
      \textsc{VarInit}
      \begin{prooftree}%\label{eq:typerule_varinit}
        \Hypo{\Gamma \vdash e : \tau'}
        \Hypo{\tau' \leq \tau}
        \Infer2{\Gamma \vdash \tau \text{ } x = e : \Gamma \cup \{x = \tau \}}
      \end{prooftree}
    \end{center}
    \begin{center}
      \textsc{Paren}
      \begin{prooftree}%\label{eq:typerule_paren}
        \Hypo{\Gamma \vdash e : \tau}
        \Infer1{\Gamma \vdash (e) : \tau}
      \end{prooftree}
      \textsc{Assign}
      \begin{prooftree}%\label{eq:typerule_assign}
        \Hypo{\Gamma \vdash x : \tau}
        \Hypo{\Gamma \vdash e : \tau'}
        \Hypo{\tau' \leq \tau}
        \Infer3{\Gamma \vdash x = e : \tau}
      \end{prooftree}
    \end{center}
    %-- container and particle access -------------------------------------
    \begin{center}
      \textsc{VecAcc}
      \begin{prooftree}%\label{eq:typerule_vecacc}
        \Hypo{\Gamma \vdash v : \mathbb{V}\langle\tau\rangle}
        \Hypo{\Gamma \vdash i : \mathbb{Z}}
        \Hypo{i \geq 0}
        \Infer3{\Gamma \vdash v[i] : \tau}
      \end{prooftree}
      \textsc{MatAcc}
      \begin{prooftree}%\label{eq:typerule_matacc}
        \Hypo{\Gamma \vdash m : \mathbb{M}\langle\tau\rangle}
        \Hypo{\Gamma \vdash i,j : \mathbb{Z}}
        \Hypo{i,j \geq 0}
        \Infer3{\Gamma \vdash m[i][j] : \tau}
      \end{prooftree}
    \end{center}
    \begin{center}
      \textsc{PartScaAcc}
      \begin{prooftree}%\label{eq:typerule_partscaacc}
        \Hypo{\Gamma \vdash p : \mathbb{P}}
        \Hypo{\Gamma \vdash f : \mathcal{E}\langle \tau, 1 \rangle}
        \Infer2{\Gamma \vdash \arrowOp{p}{f} : \tau}
      \end{prooftree}
      \textsc{PartVecAcc}
      \begin{prooftree}\label{eq:typerule_partvecacc}
        \Hypo{\Gamma \vdash p : \mathbb{P}}
        \Hypo{\Gamma \vdash f : \mathcal{E}\langle \tau, n \rangle, n \geq 2}
        \Infer2{\Gamma \vdash \arrowOp{p}{f} : \mathbb{V}\langle\tau\rangle}
      \end{prooftree}
    \end{center}
    %-- unary and binary operations ---------------------------------------
    \begin{center}
      \textsc{Unary}
      \begin{prooftree}%\label{eq:typerule_unary}
        \Hypo{\Gamma \vdash e : \tau}
        \Hypo{\tau_\ominus(\tau) \neq \bot}
        \Infer2{\Gamma \vdash \ominus\,e : \tau_\ominus(\tau)}
      \end{prooftree}
      \textsc{BinLog}
      \begin{prooftree}%\label{eq:typerule_binlog}
        \Hypo{\Gamma \vdash e_1 : \mathbb{B}}
        \Hypo{\Gamma \vdash e_2 : \mathbb{B}}
        \Infer2{\Gamma \vdash e_1 \otimes_{log} e_2 : \mathbb{B}}
      \end{prooftree}
	  \end{center}
    \begin{center}
      \textsc{BinRel}
      \begin{prooftree}%\label{eq:typerule_binrel}
        \Hypo{\Gamma \vdash e_1 : \tau_1}
        \Hypo{\Gamma \vdash e_2 : \tau_2}
         \Hypo{\tau_\otimes(\tau_1,\tau_2) \neq \bot}
        \Infer3{\Gamma \vdash e_1 \otimes_{rel} e_2 : \mathbb{B}}
      \end{prooftree}
      \textsc{BinAri}
      \begin{prooftree}%\label{eq:typerule_binarith}
        \Hypo{\Gamma \vdash e_1 : \tau_1}
        \Hypo{\Gamma \vdash e_2 : \tau_2}
        \Hypo{\tau_\otimes(\tau_1,\tau_2) \neq \bot}
        \Infer3{\Gamma \vdash e_1 \otimes_{arith} e_2 : \tau_\otimes(\tau_1, \tau_2)}
      \end{prooftree}
    \end{center}
    \begin{center}
      \textsc{ErrUnary}
      \begin{prooftree}
        \Hypo{\Gamma \vdash e : \tau \quad \tau_{\ominus}(\tau) = \bot }
        \Infer1{\Gamma \vdash \ominus(e) : \mathbb{E}}
      \end{prooftree}
      \qquad
      \textsc{ErrBin}
      \begin{prooftree}
        \Hypo{\Gamma \vdash e_1 : \tau_1}
        \Hypo{\Gamma \vdash e_2 : \tau_2 \quad \tau_{\otimes}(\tau_1, \tau_2) = \bot}
        \Infer2{\Gamma \vdash e_1 \otimes e_2 : \mathbb{E}}
      \end{prooftree}%
    \end{center}%
    %-- bottom box --------------------------------------------------------
    \begin{center}%
      \begin{tcolorbox}[enhanced,colback=white,left=0pt,right=0pt,top=0pt,boxsep=0pt,boxrule=0pt,toprule=1pt,colframe=white!60!black,rightrule=1pt,arc=0pt,outer arc=0pt,leftrule=1pt,bottomrule=1pt,width=.95\textwidth,after skip=0pt,before skip=0pt]
      \footnotesize
        \begin{minipage}{\textwidth}
         \begin{equation*}
           \ominus         \in \{-, !, \sqrt{\phantom{e}}\},\;
           \otimes_{arith} \in \{ +, -, *, /, a^b \},\;
           \otimes_{log}   \in \{ \&\&, || \},\;
           \otimes_{rel}   \in \{ ==, !=, <, >, <=, >= \}
         \end{equation*}
        \end{minipage}
        \begin{minipage}{\textwidth}
          \begin{equation*}
            \revii{\mathbb{B} = \mathit{Boolean}},\;
            \mathbb{Z} = \mathit{Integer},\;
            \mathbb{R} = \mathit{Real},\;
            \mathbb{P} = \mathit{Particle},\;
            \mathbb{E} = \mathit{Error}
          \end{equation*}
        \end{minipage}
        \begin{minipage}{\textwidth}
          \begin{equation*}
            \mathbb{V} = \mathit{Vector},\;
            \mathbb{M} = \mathit{Matrix},\;
            \mathcal{E} = \mathit{Field/Property},\;
          \end{equation*}%
        \end{minipage}%
      \end{tcolorbox}%
    \end{center}%
    % ---------------------------------------------------------------------
    \caption{Type rules for expressions in PPME.}%
    \label{fig:type_rules}%
  \end{figure}%

  Figure~\ref{fig:type_rules} shows the type rules for expressions implemented in
  PPME. The type of a variable is given by the typing environment $\Gamma$ (rule
  \textsc{Var}). A variable declaration adds a new entry to the typing environment
  (rules \textsc{VarDecl} and \textsc{VarInit}). Type rules for unary
  (\textsc{Unary}) and binary operations (\textsc{Bin*}) can be defined with a
  general scheme, where the derived type information depends on the operation
  ($\ominus$ or $\otimes$) and the types of the operand(s). This also simplifies the
  implementation of the type system and makes it extensible.%\jc{Since you do a detailed intro to the notation, should you introduce $\vdash$?}
  The type inference for unary operators $\ominus\in\{-,\sqrt{\phantom{x}},!\}$ can
  be summarized as follows: The logical not $!$ can be applied only to boolean
  arguments $e : \bool$ and its result is boolean as well. The unary minus is
  applicable for numerical expressions with $\tau \in \{ \integer, \real \}$ and will
  not change their types. Similarly, the square root operator can be applied to
  arguments of numerical type and the result will be a real number (or a runtime exception
  if the result would be a complex number, which is not included in the current static type system).
  The more detailed type-inference tables for binary arithmetic expressions
  $\tau_\otimes$ can be found in
  Figure~\ref{tab:typesystem_binary_operation_exponentiation}. In the tables,
  abbreviated forms for the types are used, where $\mathcal{E}$
  denotes a particle property or discretized data from a field.
  %in the PDE loop. We use this information in the code generator to derive the according statements that
  %were needed to be added manually by the user in PPML (using things like "sca_fields" etc.).
  % But I agree, for the type analysis, we probably do not need to make this distinction.
  The integers $n$ are $m$ denote the size of the data (i.e., the number of
  elements in the vector). Additionally, $\uparrow(\tau_1, \tau_2)$
  denotes the least common super-type of $\tau_1$ and $\tau_2$. Note that, for the
  sake of brevity, we did not include the inference tables of the remaining
  expressions. If $\tau_\otimes$ is undefined it is denoted by $\bot$.
  Moreover, the rules that end on \textsc{Acc} (\textsc{*Acc}) define the type
  inference for scalar and vector access to particle properties, which is a core task
  of the system.

  % -- type inference tables ------------------------------------------------
  \begin{figure}[tp]
    \begin{minipage}{\textwidth}
      \centering
      \scriptsize
      \renewcommand*{\thefootnote}{\fnsymbol{footnote}}
      \begin{tabular*}{\textwidth}{@{\extracolsep{\fill} } c|cccc}
        \toprule
        \parbox[b][][b]{1.5cm}{$\bm{\tau_{+|-} (\tau_1, \tau_2)}$}
          & \integer & \real & \vect{X} %& \field{X,n}
          & \prop{X,n}
        \\
        \midrule
        \integer & \integer & \real & \vect{\uparrow(\integer,X)} %& \field{\uparrow(\integer,X), n}
                             & \prop{\uparrow(\integer,X), n}
        \\
        \real & \real & \real & \vect{\uparrow(\real,X)} %& \field{\uparrow(\real,X), n}
                              & \prop{\uparrow(\real,X), n}
        \\
        \vect{Y} & \vect{\uparrow(Y, \integer)} & \vect{\uparrow(Y, \real)} & \vect{\uparrow(Y, X)} & $\bot$ %& ---
        \\
  %      \field{Y, m} & \field{\uparrow(Y, \integer),m} & \field{\uparrow(Y, \real),m} & --- & \field{\uparrow(Y, X), n}\footnotemark[2] & ---
    %    \\
        \prop{Y, m} & \prop{\uparrow(Y, \integer),m} & \prop{\uparrow(Y, \real),m} & $\bot$ %& ---
          & \prop{\uparrow(Y, X), n}\footnotemark[2]
        \\
        \bottomrule
      \end{tabular*}
      \raggedright \footnotemark[2] if $n = m$
      \vspace{2pt}
    \end{minipage}
    \vspace{5pt}
    \begin{minipage}{\linewidth}
      \centering
      \scriptsize
      \begin{tabular*}{\textwidth}{@{\extracolsep{\fill} } c|cccc} % chkTex  -2
        \toprule
 %       $\bm{\tau_{*}(\tau_1, \tau_2)}$ & \integer & \real & \vect{X} & \field{X,n} & \prop{X,n}
        $\bm{\tau_{*}(\tau_1, \tau_2)}$ & \integer & \real & \vect{X}  & \prop{X,n}
        \\
        \midrule
        \integer & \integer & \real & \vect{\uparrow(\integer, X)} & \prop{\uparrow(\integer, X),n}
      %  \integer & \integer & \real & \vect{\uparrow(\integer, X)} & \field{\uparrow(\integer, X),n} & \prop{\uparrow(\integer, X),n}
        \\
        \real & \real & \real & \vect{\uparrow(\real, X)}  & \prop{\uparrow(\real, X),n}
 %       \real & \real & \real & \vect{\uparrow(\real, X)} & \field{\uparrow(\real, X),n} & \prop{\uparrow(\real, X),n}
        \\
         \vect{Y} & \vect{\uparrow(Y, \integer)} & \vect{\uparrow(Y, \real)} & $\bot$ & $\bot$
%      \vect{Y} & \vect{\uparrow(Y, \integer)} & \vect{\uparrow(Y, \real)} & --- & --- & ---
        \\
%        \field{Y,m} & \field{\uparrow(Y, \integer),m} & \field{\uparrow(Y, \real),m} & --- & --- & ---
 %       \\
        \prop{Y,m} & \prop{\uparrow(Y, \integer),m} & \prop{\uparrow(Y, \real),m} & $\bot$ & $\bot$
%         \prop{Y,m} & \prop{\uparrow(Y, \integer),m} & \prop{\uparrow(Y, \real),m} & --- & --- & ---
        \\
        \bottomrule
      \end{tabular*}
    \end{minipage}
    \vspace{5pt}
    \begin{minipage}{\linewidth}
      \scriptsize
      \begin{tabular*}{\textwidth}{@{\extracolsep{\fill} } c|cccc} % chkTex  -2
        \toprule
%        $\bm{\tau_{/}(\tau_1, \tau_2)}$ & \integer & \real & \vect{X} & \field{X,n} & \prop{X,n}
        $\bm{\tau_{/}(\tau_1, \tau_2)}$ & \integer & \real & \vect{X} & \prop{X,n}
        \\
        \midrule
%        \integer & \real & \real & \vect{\tau_{/} (I, X)} & \field{\tau_{/} (I, X),n} & \prop{\tau_{/} (I, X),n}
        \integer & \real & \real & \vect{\tau_{/} (I, X)}  & \prop{\tau_{/} (I, X),n}
        \\
%        \real & \real & \real & \vect{\tau_{/} (R, X)} & \field{\tau_{/} (R, X),n} & \prop{\tau_{/} (R, X),n}
        \real & \real & \real & \vect{\tau_{/} (R, X)} & \prop{\tau_{/} (R, X),n}
        \\
%        \vect{Y} & \vect{\tau_{/} (Y, \real)} & \vect{\tau_{/} (Y, \real)} & --- & --- & ---
        \vect{Y} & \vect{\tau_{/} (Y, \real)} & \vect{\tau_{/} (Y, \real)} & $\bot$ & $\bot$
        \\
 %       \field{Y,m} & \field{\tau_{/} (Y, \real),m} & \field{\tau_{/} (Y, \real),m} & --- & --- & ---
  %      \\
 %       \prop{Y,m} & \prop{\tau_{/} (Y, \real),m} & \prop{\tau_{/} (Y, \real),m} & --- & --- & ---
         \prop{Y,m} & \prop{\tau_{/} (Y, \real),m} & \prop{\tau_{/} (Y, \real),m} & $\bot$ & $\bot$
        \\
        \bottomrule
      \end{tabular*}
    \end{minipage}
    \vspace{5pt}
    \begin{minipage}{\linewidth}
     \scriptsize
      \begin{tabular*}{\textwidth}{@{\extracolsep{\fill} } c|cccc} % chkTex  -2
        \toprule
%        $\bm{\tau_{a^b}(\tau_1, \tau_2)}$ & \integer & \real & \vect{X} & \field{X,n} & \prop{X,n}
        $\bm{\tau_{a^b}(\tau_1, \tau_2)}$ & \integer & \real & \vect{X} & \prop{X,n}
        \\
        \midrule
%        \integer & \integer & \real & --- & --- & ---
        \integer & \integer & \real & $\bot$ & $\bot$
        \\
%        \real & \real & \real & --- & --- & ---
        \real & \real & \real & $\bot$ & $\bot$
        \\
%        \vect{Y} & \vect{\tau_{a^b} (Y, I)} & \vect{\tau_{a^b} (Y, R)} & --- & --- & ---
        \vect{Y} & \vect{\tau_{a^b} (Y, I)} & \vect{\tau_{a^b} (Y, R)} & $\bot$ & $\bot$
        \\
      %  \field{Y,m} & \field{\tau_{a^b} (Y, \real),m} & \field{\tau_{a^b} (Y, \real),m} & --- & --- & ---
       % \\
%        \prop{Y,m} & \prop{\tau_{a^b} (Y, \real),m} & \prop{\tau_{a^b} (Y, \real),m} & --- & --- & ---
        \prop{Y,m} & \prop{\tau_{a^b} (Y, \real),m} & \prop{\tau_{a^b} (Y, \real),m} & $\bot$ & $\bot$
        \\
        \bottomrule
      \end{tabular*}%
    \end{minipage}%
    \caption{Type inference tables for binary operations $\otimes\in\{+,-\}$,  $\otimes = *$, $\otimes = /$, and exponentiation $a^b$.}
    \label{tab:typesystem_binary_operation_exponentiation} % chkTex 24
  \end{figure}
  % -------------------------------------------------------------------------

  \paragraph{Type Errors} % (fold)
  \label{par:typing_errors}

    All expressions matching one of the rules presented above are
    \emph{well-formed}, and their type can be inferred by the system. However, it
    may occur that the user enters a faulty expression for which no type inference is
    possible, yielding a \emph{typing error}. In this case, PPME has to communicate
    the error to the developer and provide meaningful information on where the error
    is located.
    To formalize this, we introduce an \emph{error type} $\mathbb{E}$ used as a
    result for non-well-formed expressions~\cite{Plotkin1981,Plotkin2004}.
    There are different causes for typing errors, e.g., incompatible types or
    undefined behavior. For instance, the exponentiation of a scalar with a string is
    not a meaningful mathematical operation and should yield a typing error in the
    corresponding expression. Error detection is not limited to arithmetic
    expressions, but covers domain-specific concepts as well.
    Furthermore, errors might be propagated, invalidating the parenting expression. To
    support this in the type system, we add two extra rules \textsc{ErrUnary} and
    \textsc{ErrBin}. A typing error $\mathbb{E}$ occurs when type resolution fails,
    i.e., if $\tau_\ominus(\tau')$ and $\tau_\otimes(\tau_1, \tau_2)$ are undefined.
    %
  % paragraph typing_errors (end)

\subsection{Dimension Annotations} % (fold)
\label{sub:dimension_annotations}

  Adding the notion of measurement units to a programming language benefits software
  developers in many ways, especially in checking the physical consistency of equations. Verifying dimensional
  integrity prevents errors in expressions that may be hard to detect
  otherwise. The language and type system extension for dimension annotations hence adds an
  additional level of analysis to detect inconsistencies at compile-time.
  Early work by \cite{Karr1978} presented a \enquote{units calculus}, a method
  to manage relationships and conversions of units, to be incorporated in programming
  languages. In \cite{Cmelik1988} and \cite{Umrigar1994} the authors
  extended the idea of measurement units to general \emph{dimensional analysis},
  covering dimensional classes of units, e.g., length or mass, meaning that
  quantities with the same dimension but different units differ only by a conversion factor.
  Dimensional analysis fits neatly into the concept of type inference in
  functional languages, establishing the base for units and dimensions in functional
  languages~\cite{Wand1991,Kennedy1994,Kennedy1997,Hayes1995}.
  %

%\paragraph{Dimension Declaration and Specification} % (fold)
%  \label{dimension_declaration_and_specification}
In PPME, we consider dimensions and units as additional annotations to types and expressions
that are processed by an extended type system, with $\mathcal{I}$ the set of
dimensions supported by this system.
%
%The developer defines the elementary dimensions, derived
%dimensions can be defined for readability.     %
Dimensions without specification, such as length $l$, mass $m$, or time $t$, are called
a \emph{fundamental}. We denote fundamental dimensions with $\check{\delta}$
and the set of fundamental dimensions as $\check{\mathcal{I}}\subseteq{\mathcal{I}}$. Additional
\emph{derived dimensions} $\delta$, such as acceleration or force, can be composed from others
by means of multiplication and exponentiation, e.g., for acceleration $a = l \cdot t^{-2}$.
While all derived dimensions can be composed from fundamental dimensions only, PPME also allows definitions from other derived dimensions to simplify notation. This is described by the following grammar:
%The dimension calculus in PPME works uniformly over fundamental, derived, and
%computed dimensions. Therefore, we introduce the notion of a \emph{dimension
%specification} $\delta \in \mathcal{I}$  of the form
%
  \[ \delta ::= \check{\delta} ~|~ \delta_1 \cdot \delta_2 ~|~ \delta^n \, , \]
where $\hat{\delta} \in \hat{\mathcal{I}}$, $n \in \mathbb{Z}$, and derived dimensions
$\delta_1$, $\delta_2$, and $\delta$.
%
% SK: NOTE: Above grammar is compact: \delta is used as nonterminal and
% element of the language. Alternatively, we could use
% \[ d::= \check{\delta} ~|~ d_1 \cdot d_2 ~|~ d^n\] where d is used as a distinguished
% nonterminal. Furthermore the grammar does not consider just reusing dimensions
% that were already defined so that terms, e.g., for acceleration may repeat.
%
To make dimensions comparable, they are represented
in \emph{base form}, i.e., as a combination of $k>{0}$ fundamental dimensions $\check{\delta}_i$ raised to some integer
exponent $n_i$ where each $\check{\delta}_i$ occurs at most once.
A base form of $\delta$ can be constructed by recursively replacing all derived dimensions
in $\delta$ by their declaration, and grouping all occurrences of the same fundamental dimension
$\check{\delta}_i$ in a single equivalent power representation $\check{\delta}_i^{n_i}$.
We denote the expansion and base form of $\delta$ by $\lceil \delta \rceil$ with
  \[ \lceil \delta \rceil := \{ \check{\delta}_1^{n_1}, \check{\delta}_2^{n_2}, \dots, \check{\delta}_k^{n_k} \} \, . \]
Based on this definition, two dimensions $\delta_1$ and $\delta_2$ \emph{match} if $\lceil \delta_1 \rceil
= \lceil \delta_2 \rceil$, denoted by $\delta_1 \equiv \delta_2$,.

  % paragraph dimension_declaration_and_specification (end)

 % \paragraph{Dimension Type Rules and Errors} % (fold)
%  \label{par:dimension_type_rules_and_errors}

    Dimensions can be easily integrated into the PPME type system by extending it
    with dimension-specific rules and retaining the original inference mechanism.
%    We utilize the type system engine for checking and infering dimensions. Thus, the
 %   formal type system presented in \ref{sec:type_rules} is extended with rules for
 %  the unit calculus. The type inference is deferred to the original type system.
%
Given a type $\tau$ and a dimension $\delta$, we denote the
\emph{annotated type} by $\hat{\tau} = [\tau; \delta]$. In particular, any type
$\tau$ can be annotated with the \emph{empty dimension} $\emptyset$ without
changing semantics by using $e : [\tau; \emptyset]$ instead of $e : \tau$. The annotation of
metadata to types or literals is denoted by curly braces, i.e., $\tau\{\delta\} :
[\tau; \delta]$, and $e\{\delta\} : [\tau;\delta]$ instead of $e : \tau$.
Moreover, for dimension inference, we use a notation similar to the
type inference table in Figure~\ref{tab:typesystem_binary_operation_exponentiation}.
For instance, $\delta =
\mathcal{I}_\otimes(\delta_1,\delta_2)$ denotes that $\delta$ is inferred from the
operand dimensions $\delta_1$ and $\delta_2$ and the operation $\otimes$.

% JC: Next as separate paragraph - (issue 13)
Finally, the original type rules shown in Figure~\ref{fig:type_rules} need to be adapted.
As this adaptation is mostly straightforward, we only show the most relevant rules in
Figure~\ref{fig:dimension_rules}.
%%%% SK: explanations of the rules can potentially be omitted
The rules for handling variable references
(\textsc{Var}) and variable declarations (\textsc{VarDecl} and \textsc{VarInit})
have been expanded with annotated dimensions.  Assignment expressions
(\textsc{Assign}) now take the annotated dimension into account.  The general
scheme for unary (\textsc{Unary}) and binary operations
(\textsc{BinOp}) is, likewise, extended for annotated types.
Besides type inference ($\tau_\ominus$ and $\tau_\otimes$), dimensions are inferred through
$\mathcal{I}_\ominus$ and $\mathcal{I}_\otimes$. The rule \textsc{ErrDim} exemplary
shows the additional potential for error detection introduced by dimensions: even if types match,
a dimension error is still detectable.

    %-- dimension inference rules ---------------------------------------------------
    \begin{figure}[t]
      \footnotesize
      \begin{center}
         \textsc{Var}
        \begin{prooftree}[separation=1em]
          \Hypo{\Gamma(v) = [\tau; \delta]}
          \Infer1{\Gamma \vdash v : [\tau; \delta]}
        \end{prooftree}
        \textsc{VarDecl}
        \begin{prooftree}[separation=1em]
          \Hypo{\phantom{\Gamma}}
          \Infer1{\Gamma \vdash \tau\{\delta\}\, x : \Gamma \cup\{ x =  [\tau; \delta]\}}
        \end{prooftree}
        \textsc{VarInit}
        \begin{prooftree}[separation=1em]
          \Hypo{\Gamma \vdash e : [\tau'; \delta']}
          \Hypo{\tau' \leq \tau}
          \Hypo{\delta' \equiv \delta}
          \Infer3{\Gamma \vdash \tau\{\delta\}\, x = e: \Gamma \cup\{ x =  [\tau; \delta]\}}
        \end{prooftree}
      \end{center}
      \begin{center}
        \textsc{ASSIGN}
        \begin{prooftree}[separation=1em]
          \Hypo{\Gamma \vdash x : [\tau; \delta]}
          \Hypo{\Gamma \vdash e : [\tau'; \delta']}
          \Hypo{\tau' \leq \tau}
          \Hypo{\delta' \equiv \delta}
          \Infer4{\Gamma \vdash x = e : [\tau; \delta]}
        \end{prooftree}
        \textsc{Unary}
        \begin{prooftree}[separation=1em]
          \Hypo{\Gamma \vdash e : [\tau; \delta]}
          \Hypo{\tau_\ominus(\tau) \neq \bot}
          \Hypo{\mathcal{I}_\ominus(\delta) \neq \bot}
          \Infer3{\Gamma \vdash \ominus\,e : [\tau_\ominus(\tau); \mathcal{I}_\ominus(\delta)]}
        \end{prooftree}
        \end{center}
        \begin{center}
        \textsc{BinOp}
        \begin{prooftree}[separation=1em]
          \Hypo{\Gamma \vdash e_1 : [\tau_1; \delta_1]}
          \Hypo{\Gamma \vdash e_2 : [\tau_2; \delta_2]}
          \Hypo{\tau_\otimes(\tau_1,\tau_2) \neq \bot}
          \Hypo{\mathcal{I}_\otimes(\delta_1,\delta_2) \neq \bot}
          \Infer4{\Gamma \vdash e_1 \otimes e_2 : [\tau_\otimes(\tau_1, \tau_2); \mathcal{I}_\otimes(\delta_1, \delta_2)]}
        \end{prooftree}
      \end{center}
      \begin{center}
        \textsc{ErrDim}
        \begin{prooftree}[separation=1em]
          \Hypo{\Gamma \vdash e_1 : [\tau_1; \delta_1]}
          \Hypo{\Gamma \vdash e_2 : [\tau_2; \delta_2]}
          \Hypo{\tau_\otimes(\tau_1,\tau_2) \neq \bot} % we give preference to type errors ...
          \Hypo{\mathcal{I}_\otimes(\delta_1, \delta_2) = \bot}
          \Infer4{\Gamma \vdash e_1 \otimes e_2 : \mathbb{E}}
        \end{prooftree}
      \end{center}
      \caption{Type rules for dimension-annotated expressions in PPME.}
      \label{fig:dimension_rules}
    \end{figure}
    %--------------------------------------------------------------------------------

     %
    \begin{figure}
      \centering
      \ssmall
      \begin{prooftree}[separation=1em]
        \Hypo{\color{black}\Gamma \vdash \arrowOp{p}{v} : [\real, v]}
        \Hypo{\Gamma \vdash 0.5 : [\real, \emptyset]}
        \Hypo{\Gamma \vdash \arrowOp{p}{a} : [\real, a]}
        \Hypo{\Gamma \vdash \arrowOp{p}{F} : [\real, m \cdot a]}
        \Hypo{\Gamma \vdash \mathit{mass} : [\real, m]}
        \Infer2[(1)]{\Gamma \vdash \arrowOp{p}{F} / \mathit{mass} : [\real, a]}
        \Infer2[(2)]{\Gamma \vdash \arrowOp{p}{a} + \arrowOp{p}{F} / \mathit{mass} : [\real, a]}
        \Infer2[(3)]{\Gamma \vdash 0.5 * (\arrowOp{p}{a} + \arrowOp{p}{F} / \mathit{mass}) : [\real, a]}
        \Hypo{\Gamma \vdash {\mathit{delta}\_t}^2 : [\real, t^2]}
        \Infer2[(4)]{\Gamma \vdash 0.5 * (\arrowOp{p}{a} + \arrowOp{p}{F} / \mathit{mass}) * {\mathit{delta}\_t}^2: [\real, a \cdot t^2]}
        \Alter{\color{black}}
        \Infer2[(E)]{\Gamma \vdash \arrowOp{p}{v} + 0.5 * (\arrowOp{p}{a} + \arrowOp{p}{F} / \mathit{mass}) * {\mathit{delta}\_t}^2 : \mathbb{E}}
        \Alter{\color{red!90!black}}
      \end{prooftree}
      \begin{tcolorbox}[enhanced,colback=white,left=0pt,right=0pt,top=2pt,bottom=2pt,boxsep=0pt,boxrule=0pt,toprule=1pt,colframe=white!60!black,rightrule=1pt,arc=0pt,outer arc=0pt,leftrule=1pt,bottomrule=1pt,width=.95\textwidth,after skip=0pt,before skip=2pt]
        \begin{center}
          $(1)\, \tau_{/}(\real,\real) = \real, \mathcal{I}_{/}(m \cdot a, m) = a$
          \qquad
          $(2)\, \tau_{+}(\real,\real) = \real, \mathcal{I}_{+}(a, a) = a$
         \qquad
          $(3)\, \tau_{*}(\real,\real) = \real, \mathcal{I}_{*}(\emptyset, a) = a$
          \\
          $(4)\, \tau_{*}(\real,\real) = \real, \mathcal{I}_{*}(a, t^2) = a \cdot t^2$
          \qquad
          ${\color{red!90!black}(E)}\, \tau_{+}(\real,\real) = \real, {\color{red!90!black}\mathcal{I}_{+}(v, a \cdot t^2) = \bot}$
        \end{center}
      \end{tcolorbox}%
      \caption{
        Example deduction using the extended type system.
        \revii{The applied type inference rule $\tau_\otimes$ and dimension inference rule $\mathcal{I}_\otimes$ for
        respective steps (1) to (4), and (E) are shown in the box. The failing dimension inference rule is marked in red.}
      }
      \label{fig:type_deduction}
    \end{figure}
    As an example for applying the type system, consider the following PPME expression
    that has been modified from the Lennard-Jones case study discussed in
    Section~\ref{sub:case_studies}:
     \[ \arrowOp{p}{v} + 0.5 * (\arrowOp{p}{a} + \arrowOp{p}{F} / \mathit{mass}) * {\mathit{delta}\_t}^{\color{red!90!black}2} \, . \]
    Here, $p$ refers to a particle whose properties $v$ (velocity) and $a$ (acceleration),
    and the force field $F$ are used together with the free variables $\mathit{mass}$ and $\mathit{delta}\_t$ (time)
    to compute an update in a larger simulation. We introduced a small error into the expression: $\mathit{delta}\_t$ has an
    exponent of $2$ instead of $1$. Since this does not have an impact on the overall type of the expression ($\mathit{\real}$),
    a conventional type system cannot detect this error. However, using dimensions, the problem becomes discoverable, as shown
    by the deduction depicted in Figure~\ref{fig:type_deduction}. In step $(1)$, the type and dimension of the subexpression
    $\arrowOp{p}{F} / \mathit{mass}$ is deduced from its compartments ($[\real, a]$).
    Step $(2)$ deduces the type of the enclosing addition \revii{of $\arrowOp{p}{a}$ and $\arrowOp{p}{F}/mass$}
    (again $[\real, a]$).
    In step $(3)$, the type remains the same because of \revii{multiplication with the constant 0.5}.
    Step $(4)$ computes the type of the multiplication with ${\mathit{delta}\_t}^2$ as $[\real, a \cdot t^2]$. Finally, in step
    $(E)$, the type system discovers that the outermost sum is infeasible since  $[\real, v]$ and $[\real, a \cdot t^2]$ are incompatible,
    deducing error type $\mathbb{E}$. In contrast, if the expression would have been correct, the calculus would have
    derived $[\real, v]$ as the overall type.
     % paragraph dimension_type_rules_and_errors (end)

% subsection dimension_annotations (end)

% vim: set ts=2 sw=2 sts=2:
% vim: set wrap breakindent tw=85 cc=85:
% !TEX root = paper_ppme_TOMS.tex

\section{The PPM Environment: Architecture and Implementation}
\label{sec:implementation_use_cases}
%The \emph{PPM Environment} (PPME) is an IDE and DSL for developing numerical simulations using the parallel particle-mesh method.
 %It aims to reduce the knowledge gap, that is, the mismatch between domain experts and the required expertise for an efficient use of HPC resources~\cite{Sbalzarini2009}.
 %Therefore, it provides high-level abstractions and notations that are well-known to domain experts and, thus, align with the class of problems relevant in particle-mesh-based simulations.
 %
%The development process of scientific applications can be improved by domain-specific tools and languages.
The \emph{productivity} of scientific programmers can be increased by providing high-level abstractions for
computational models, such that the developer is not bothered with details of the programming language
or the underlying hardware architecture. While \emph{quality} is hard to measure, an IDE can check for common errors up-front and present the developer with meaningful warnings
and error messages. Additionally, static program analysis, paired with domain knowledge, can be used to
improve performance, accuracy, and/or efficiency of simulations.
Incorporating third-party applications allows
to reuse established tools for analysis and program transformation instead of reimplementing their features. 
\revii{Full access to the underlying language and implementation enables advanced scientific programmers to leverage their knowledge to add new language-level features and to have full control over program performance.}
All of these features are available in PPME.

\subsection{Internal Structure of PPME}
\begin{figure}[tp]
    \begin{minipage}[b]{.57\textwidth}
      \centering
      \includegraphics[width=1\textwidth]{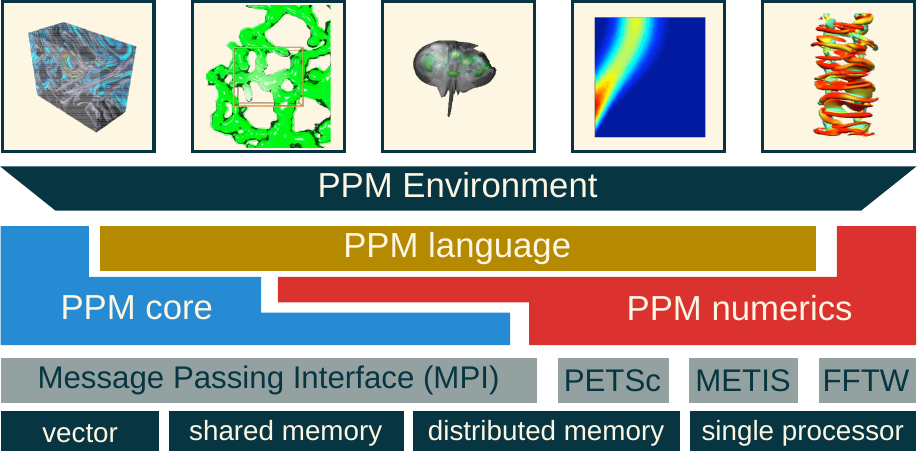}
      \caption{PPME: New access layer to the underlying PPML.}
      \label{fig:ppme-access_layer}
    \end{minipage}
    \hfill
    \begin{minipage}[b]{.40\textwidth}
      \centering
      \includegraphics[width=1\textwidth]{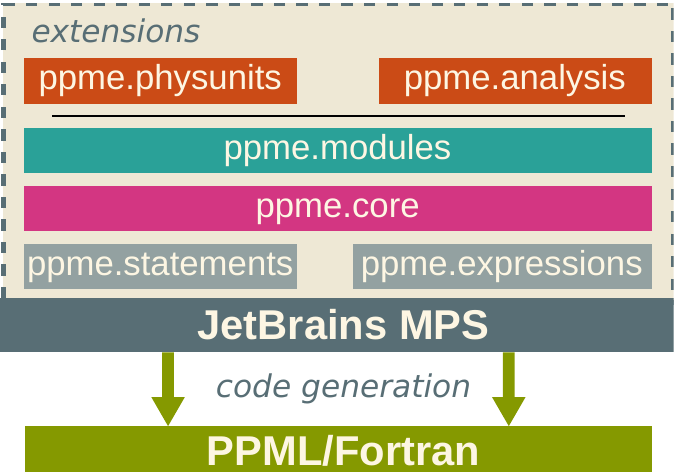}
      \caption{Modular architecture of PPME.}
      \label{fig:ppme-architecture}
    \end{minipage}
  \end{figure}

\revii{Based on the Meta Programming System (MPS)}, PPME adds an additional layer on top of the existing PPML 
stack~\cite{Sbalzarini2006a,Awile:2013a} and does not require any adaptation in the underlying framework.
It generates source code against PPML, and therefore makes use of the established workflow, using PPML
as an intermediate representation. Figure~\ref{fig:ppme-access_layer} illustrates how PPME fits between
the user program and the PPM middleware. Application developers interact with the development environment
to implement particle methods and the related configuration files. %Furthermore, the IDE provides a simpler access to the PPM library than PPML.
PPML's original purpose of hiding technical details, specific realizations, and the explicit target platform
is preserved. However, PPME offers a more consistent DSL syntax and incorporates domain-specific
elements as first-class concepts.

In PPME, the domain metamodel of Section~\ref{sec:domain-metamodel}
is organized in language packages, called \emph{solutions} in MPS. The clean separation between
different sub-languages enables a good separation of concerns. The lower layers form a base DSL with
general language constructs such as expressions, literals, and statements. These concepts are reused in the upper
layers to define domain-specific language concepts for particle-mesh methods on top of the base language.
Reusing lower layers and their extension is key in the design of PPME, enabling easier
maintenance and custom extensions for specific cases as plug-ins. 

\revii{Figure~\ref{fig:ppme-architecture} provides
a schematic view of PPME's internal language stack. The bottom layers form the main DSL while sub-languages
are used to keep the implementation modular and maintainable. At the interface to the underlying PPM library, 
MPS manages code transformation and generation. The top layer is open to new application-specific extensions, e.g., for particle-based image processing~\cite{Afshar2016}.
The languages packages of PPME (cf. Figure~\ref{fig:ppme-architecture}) are:}

\textbf{ppme.expressions}.
This package provides general expressions as can be found in most programming languages, e.g., mathematical
and logical expressions, and literals for integer and floating-point numbers. Moreover, the base types available in
PPME are defined in this package, as well as essential parts of the type system introduced in Section~\ref{sec:types_units}.
As already mentioned previously, the main purpose of the static type system is to detect illegal expressions early,
at compile time or while editing. The PPME editor therefore instantaneously analyzes the program using the type-inference rules.
When an error type $\mathbb{E}$ is derived, the editor displays an error mark and a cause-related error message, if the
cursor is hovered over the erroneous expression. 

\textbf{ppme.statements}.
The statements sub-language contains a basic set of imperatives, such as expression statements, if-else clauses, and loops.
Furthermore, variable declarations and references are part of this package.
The type system is enriched with variable support where necessary (cf. Section~\ref{sec:type_rules}).
Overall, the elements of this language are universal since they are independent of the domain they are used in.

\textbf{ppme.core}.
The core package contains most elements specific to particle methods.
It extends the solutions for expressions and statements by adding new domain-specific types, expressions, and statements.
Selected constructs of PPML are reflected in PPME while remaining consistent with the base language's concepts.
For instance, the \emph{timeloop} construct of PPML is available in this package.

\textbf{ppme.modules}.
A module in PPME is the top-level structure for client programs written in PPME. It contains the simulation code
and optional specifications for imported control parameters. A module translates to a PPML client, but the
IDE can use additional knowledge about the domain better than PPML, e.g., by referencing external control
files and inspecting the code.

\textbf{ppme.lang}.
This package is an MPS \emph{devkit} that contains the above base languages of PPME (not shown explicitly in Figure~\ref{fig:ppme-architecture}).
In MPS, devkits group interconnected languages as one unit. Hence, to get the base functionality of
PPME's language it suffices to include the devkit in an MPS project, covering all language dependencies.

In addition to these base languages, PPME provides optional solutions that add dimensions and physical units into program specifications,
and for the integration of external analysis tools. Both serve as examples for further extensions
tailored to specific use-cases:

\textbf{ppme.physunits}.
The optional physical-units integration enables developers to annotate further meta information to variables and constants.
This includes means for adding dimensions and physical unit specifications as an additional extensible
layer in the type system (cf. Section~\ref{sub:dimension_annotations}).

\textbf{ppme.analysis}
The analysis language consists of an exemplary binding of \emph{Herbie} as an external analysis tool for improving
floating-point expressions~\cite{Panchekha2015}. We elaborate a general framework enabling the access of custom
 tools in the environment. More details on this tool integration will be given later in Section~\ref{sec:analysis}.

\subsection{Code Generation}
The code-generation process in PPME involves an integrated transformation and analysis chain. This process is
sketched subsequently, followed by two concrete example transformation steps.

\subsubsection{Transformation Process}
Code generation in PPME is implemented via several model-to-model transformations
refining the program, and a final text-generation stage that produces source code in
the PPML target language. 
%
%\footnote{
\revii{Models in MPS are directed graphs with type annotations derived from the metamodel. The graphs have a distinct spanning tree, which in general corresponds to an abstract syntax tree. Model-to-model transformations map an input graph to an output graph, where the output graph may use the same or different type annotations given by the same or another metamodel. Models must adhere the structural constraints defined by their metamodels. An excerpt of the PPME metamodel is given in Figure~\ref{fig:domain-model}.}
%}
%
\revii{During the transformation process,} the internal graph-based representation of the program is
enriched with additional information that is explicitly required to generate the
output in the target language, e.g., a list of variables accessed in a loop can be
derived from the loop's body and made available explicitly for further processing. In
general, the concept of staged language processing is advantageous, for example to yield
different output representations of an input program, or for transformation chaining.
%
%\tn{elaborate this aspect in more detail with focus on LWs?} The concept of
%transformation stages allows for different fields off application. On the one hand,
%several transformations for the same model could be executed in parallel, yielding
%several representations of a single input program (e.\,g., producing customized
%versions for % several target platforms). On the other hand, transformations can be
%chained, i.\,e., the result of one transformation is used as the input of the next
%one. Since PPME aims for modular language extension and composition, the underlying
%language workbench has to support transformations introduced by different language
%extensions.

Dependencies between transformation steps are resolved automatically so that a global
transformation sequence for a given program can be computed. This allows adding new
features and further transformations to extend PPME without affecting
other components.

%Supporting PPME's goals towards language extensibility, MPS automatically resolves
%dependencies and computes a global transformation sequence for a given
%program\jc{Turn this around, as a feature of PPME thanks to MPS}. This allows adding
%new features and plug in transformations without affecting other components.

\begin{figure}[tp]
  \centering
  \includegraphics[scale=0.7]{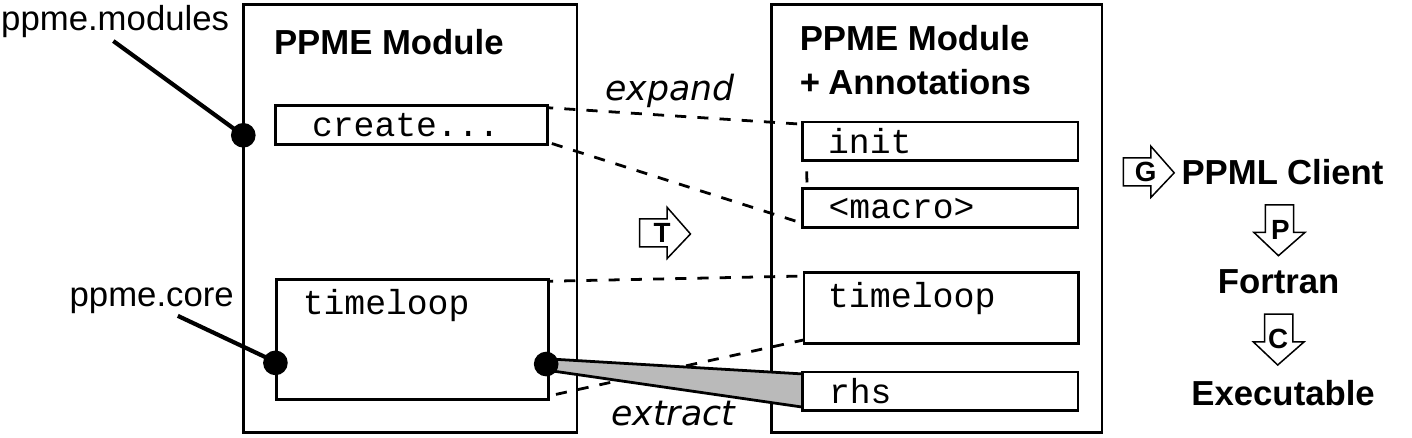}
  \caption{%
    Domain-specific abstractions in a PPME module are transformed (\textbf{T}) into
    lower-level representations. From this enriched module, PPML client code is
    generated (\textbf{G}) which is subsequently processed (\textbf{P}) to Fortran
    code and compiled (\textbf{C}) to a binary linking against the PPM Library.%
  }
  \label{fig:ppme-transformation} % chkTex 24
\end{figure}
To avoid unnecessary overhead during the generation phase, textual output is produced
in only two cases: (a) when the final output in the target language is generated, and
(b) when external tools and analyses require textual input. This restriction enables
full control of the transformation phase within MPS, taking into account the
enrichments of various transformations and results of external components, such as
Herbie and other tools. The produced source code can then be compiled using a regular
compiler. 

% JC: Next as separate paragraph - (issue 13)
The code generation process is illustrated in Figure~\ref{fig:ppme-transformation}. It starts
with a simulation program implemented using the domain metamodel introduced in
Section~\ref{sec:domain-metamodel}. The module contains domain-specific concepts such
as a \inline{timeloop} and various constructs to define particle-based
simulations. In Figure~\ref{fig:ppme-transformation}, the \inline{timeloop} statement
is analyzed in the first transformation stage and a PPML right-hand side specification
(cf. Figure~\ref{fig:grayscott-ppml}, Lines $53$--$65$) is extracted. Similarly, the
creation of particles is expanded to several initialization and macro calls in the
representation of the module, which is closer to the target language.

The majority of the model-transformations are part of the top-level package
\inline{ppme.modules}. Various mapping scripts are used to \emph{pre-process} the
input-model for collecting information required in later transformation steps.
To produce the intermediate code (i.e., PPML code), MPS' \emph{text
gen} capabilities are used (cf. \cite{MPS32TextGen}). For each language concept, a
text generation component can be specified, defining its textual representation,
e.g., printing the name of a variable (\inline{VariableReference}), or emitting the
code for a loop statement.

%\tn{describe general capabilities of LWs for target code generation}\sk{not sure, if
%we would need this}
We use several transformation scripts to prepare the text-generation phase.
Therefore, we have defined multiple intermediate models resembling the macros and
first-class language constructs in PPML to stepwise refine the input model. This
includes collecting information about the differential operators used in equations,
adding explicit discretization statements for them, creating and populating right-hand side
declarations of PPML, adding ODE declarations, managing control files, expanding
random-number initialization, deriving field and property declarations, transforming
foreach loops into their PPML counterparts, and adding PPML-specific type annotations.
Exemplarily, we discuss the construction of right-hand-side declarations.

\subsubsection{Example Transformation}
%\jc{Is it possible to list all transformations, probably with a short description,
%and then only explain the ones you explained in detail. This would help to remove
%vagueness.}
%
%%
%\begin{figure}[tp]
%  \centering
%  \includegraphics[width=0.75\textwidth]{img//ppme-replaceRandoms.pdf}
%  \caption{%
%  Program-graph transformation translating random-number expressions to variable references (red arrows).
%  }
%  \label{fig:ppme-replaceRandoms} % chkTex 24
%\end{figure}
%%
%PPME offers high-level abstractions for initializing and using random-number generators from
%within an expression by writing, e.g., \inline{random<integer>} to generate a random integer value. 
%The system transforms this into corresponding code by adding a variable-declaration to the preamble
% and replacing the random expressions by references to the declared variable.
%Figure~\ref{fig:ppme-replaceRandoms} illustrates this transformation. The expression
%is replaced by a variable reference to \inline{rnd_1} and a random value is
%assigned to this variable right before the expression statement. 

%\sk{This random-number stuff could be removed. It has too many flaws.}

\begin{figure}[t]
  \includegraphics{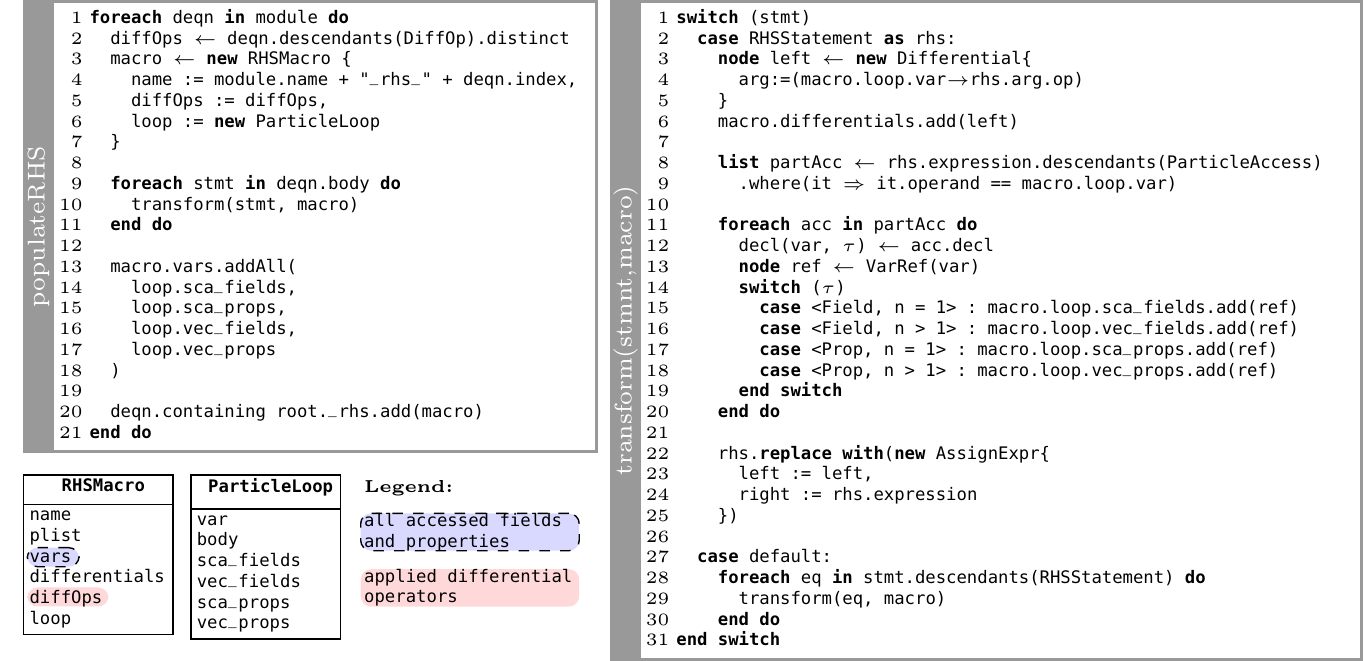}
  \caption{Excerpts from the script that composes a PPML \inline{RHSMacro} from a PPME \inline{deqn} specification.}
  \label{fig:ppme-populateRHS}
\end{figure}

The \inline{populateRHS} transformation is responsible for extracting
right-hand-side definitions from \inline{deqn} statements\revii{, which model differential equations.} \revii{In PPME, these equations can be written
directly in code.
To transform them into PPML, we use \inline{RHSMacro}s, which represent 
the right-hand side definitions in the target language and, in turn, can be transformed into PPML macro code.}
The \inline{deqn} statements are matched and extracted by the code generator using transformation scripts \revii{in MPS' built-in scripting language}.
The scripting language is statically typed and borrows ideas from object-oriented and functional programming that are well-suited for model transformations, \revii{including}
higher-order functions, and type-based selectors on trees and lists. \revii{For example, the \inline{descendants} selector visits a tree or list and collects references to nodes of a given type, like iterators in an object-oriented language. 
In addition, higher-order functions such as \inline{map} or \inline{where} allow developers to map or filter lists by applying a given (anonymous) function, like in a functional program.} 
The left panel 
of Figure~\ref{fig:ppme-populateRHS} shows the script that implements the \inline{populateRHS} transformation. 
It iterates over all blocks of \inline{deqn} statements in a
program, and derives corresponding \inline{RHSMacro}s. A key issue in this process is to identify the used
differential operators \revii{contained in a \inline{deqn} statement}  (l. 2) and the accessed variables (ll. 13--18). 
A \revii{PPML} particle-loop, evaluating the differential operators over these variables, is assembled by transforming each statement in
the \inline{deqn} \revii{via delegation to \inline{transform}} (ll. 9--11). \revii{Finally, the created macro object is added to the model root and then used for generating the PPML right-hand side.}

% JC: Next as separate paragraph - (issue 13)
The script excerpt in the right panel of Figure~\ref{fig:ppme-populateRHS} shows the body
of the recursive \inline{transform()} function. It takes a statement of a
differential equation and enriches the given macro with information. For each
differential equation statement $\frac{\partial\arrowOp{c}{f}}{\partial{t}} = e$
(\inline{stmt}), the affected particle attribute ($f$) is extracted (ll. 2--6).
Subsequently, the accessed particle fields and properties are extracted and explicitly
added to the \texttt{RHSMacro} (ll. 11--20). Finally, the differential equation is
translated to an assignment expression \revii{that replaces the original definition in the model and  is later translated to PPML statements for the generated code of the right-hand side}.

\begin{figure}[t]
\centering
\includegraphics{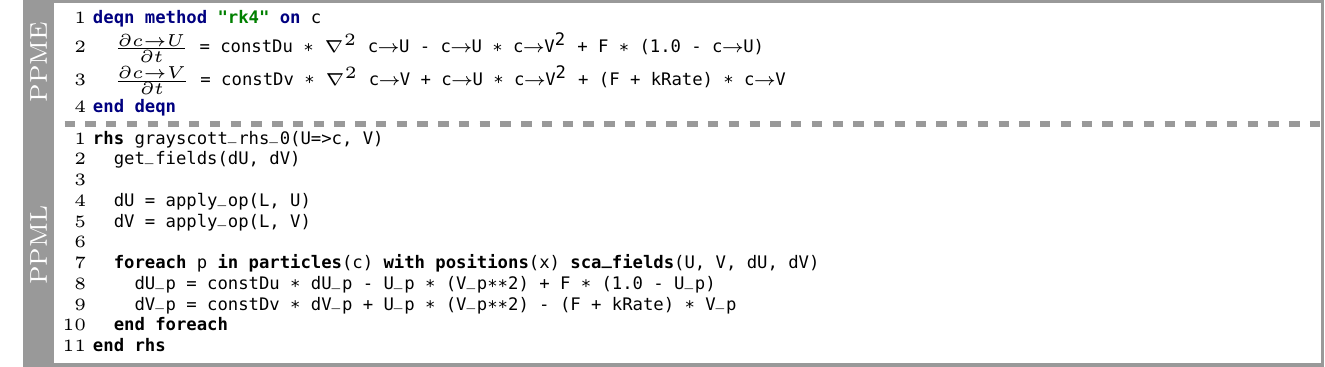}
\caption{PDE specification in PPME (top) and the generated right-hand side in PPML (bottom).}
\label{fig:ode-comparison}
\end{figure}

Figure~\ref{fig:ode-comparison} shows a comparison of the original PPME block and the
resulting PPML code for the Gray-Scott example (cf. Section~\ref{sec:ppml}). 
It integrates the governing equations discretized over the particle list $c$ using the 
4-th-order Runge-Kutta method (``rk4'') for time integration. 
In PPME, a developer conveniently defines equations over attributes of a particle
list \inline{c}. The IDE automatically extracts the required information. First, two
applications of differential operators,
\lstinline[mathescape]{$\nabla^2$c$\rightarrow$U} and
\lstinline[mathescape]{$\nabla^2$c$\rightarrow$V}, are identified. The local
variables \inline{dU} and \inline{dV} are inserted to hold the intermediate result of
applying the operators. Both the particle loop and the right-hand-side block itself
hold derived information about the accessed particle fields, \inline{U} and
\inline{V}. Furthermore, \inline{dU} and \inline{dV} are treated like other scalar
fields. Note that the access of particle list attributes (\inline{c$\rightarrow$U})
is transformed to access of particle attributes by inserting a loop over particles and accessing attributes of the
loop variable \inline{U_p}.
The example demonstrates some of the key benefits of our approach over the
original PPML code: since all required information is extracted by the PPME
compiler, redundant statements such as \inline{get_fields}, \inline{apply_op} and
\inline{sca_fields} are avoided, which leads to less code, less compile-time errors, 
and an improved readability. Readability is further improved by the PPME editor natively 
supporting basic mathematical notation, such as the Nabla operator $\nabla$ and the partial 
derivative $\partial$.
% subsection code_generation (end)

% vim: set ts=2 sw=2 sts=2:
% vim: set wrap breakindent:
% !TEX root = master.tex        --- atom

\subsection{Case Studies} % (fold)
\label{sub:case_studies}

To demonstrate the capabilities of PPME, we use the same two simulations as case
studies that were already considered for PPML \cite{Awile:2013a}. The first one, the Gray-Scott reaction-diffusion system as presented in
Section~\ref{sub:a_simple_application_example}, is an example of a simulation of a continuous
deterministic model. The second one, Lennard-Jones molecular dynamics is an example of a simulation of a 
discrete deterministic model. An N-body simulation as a third example further illustrates the initialization of particles from external data. 

\subsubsection{Gray-Scott Reaction-Diffusion System} % (fold)
\begin{figure}[tp]
    \centering
    \begin{minipage}{.49\textwidth}
      \includegraphics[scale=1.0]{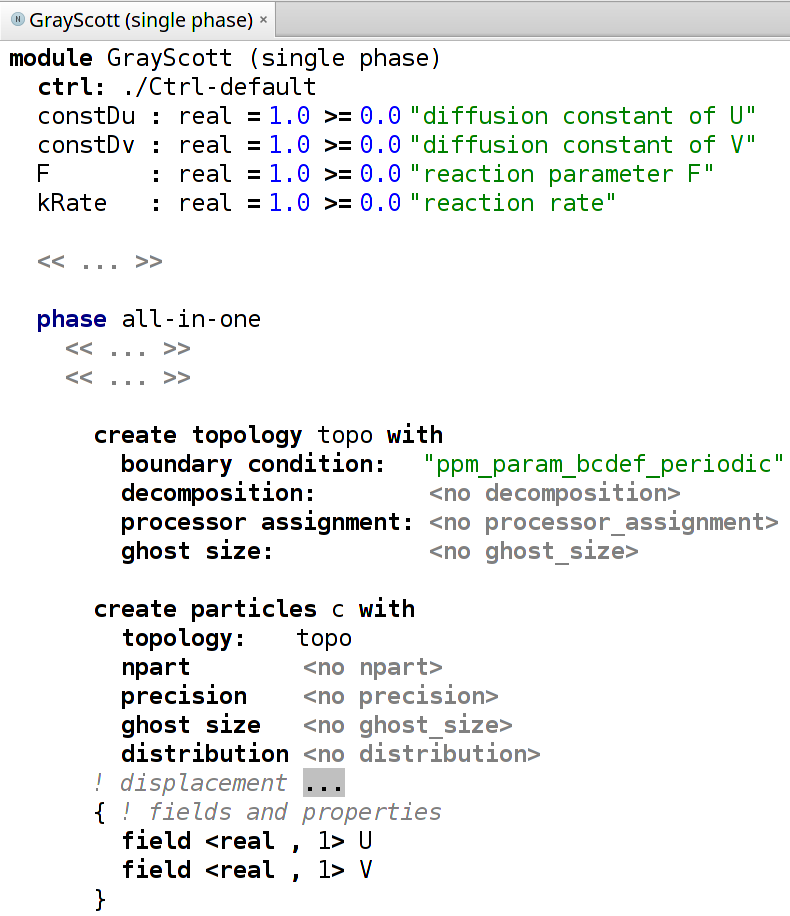}
    \end{minipage}
    \begin{minipage}{.50\textwidth}
      \includegraphics[scale=1.0]{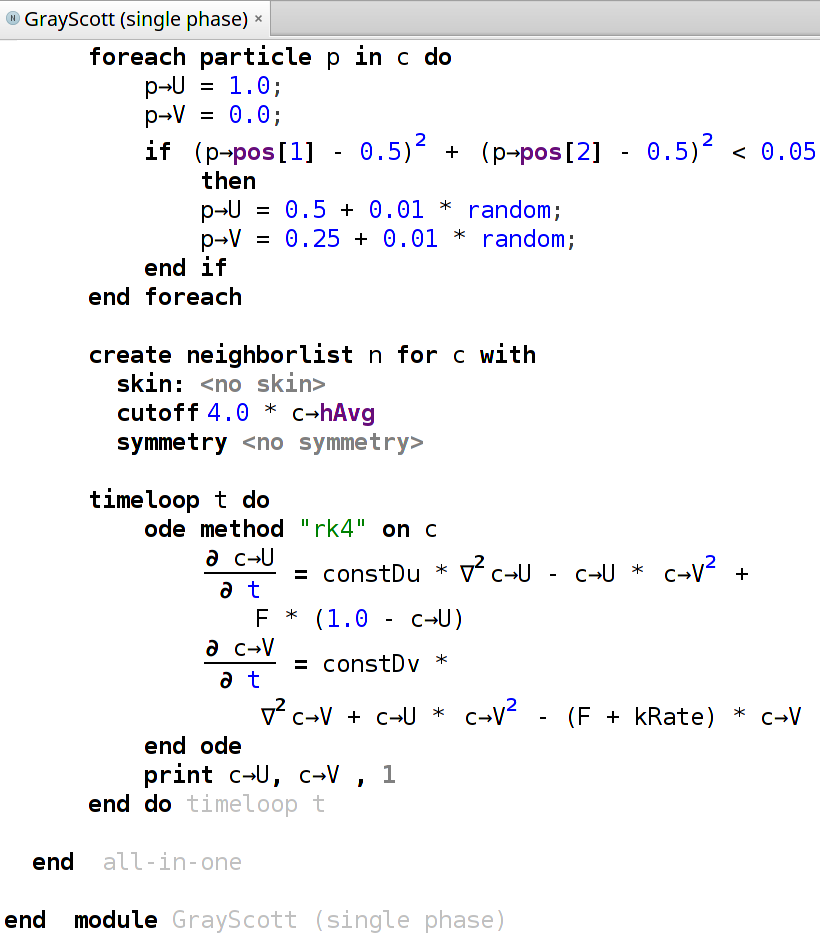}
    \end{minipage}
    \caption{Gray-Scott simulation program in PPME.}
    \label{fig:ppme-gs}
  \end{figure}

\label{par:gray_scott_reaction_diffusios_system}
  The PPME program for the Gray-Scott simulation is shown in Fig.~\ref{fig:ppme-gs}. It follows the typical structure of a particle-based simulation, starting
  with the initialization of topology and particles, followed by the simulation loop.
  The notation in PPME is concise and close to the domain idiom.
  The program starts with
  the module definition and the referenced runtime constants. At the beginning of the
  simulation, topology, particles, and neighbor lists are set up. The time steps are contained in the \inline{timeloop} and are solely defined through
  the differential equations to be solved. For the equation block, the developer has to specify the
  particle list the equations are working on, and the time-stepping method. Note that
  the continuous fields $U$ and $V$ are automatically discretized on the particle
  list during code generation. 

% paragraph gray_scott_reaction_diffusios_system (end)

\subsubsection{Lennard-Jones Molecular Dynamics} % (fold)
\label{par:lennard_jones_potential}
  Lennard-Jones is an instance of
  molecular dynamics~\cite{Frenkel2001}, an item-based simulation to study molecular 
  processes. The atoms are directly represented as particles,
  located in continuous space. Pairwise potentials between atoms define the
  continuous forces acting on them. While the basic algorithm for the simulation,
  i.e., computing pairwise interactions of particles and updating their positions
  and properties, remains the same, the exact definition of the forces is specific to
  the application. A classical force definition is given by the Lennard-Jones potential, which is suitable for describing inert gases. 
  The pairwise force between 
  atoms depends on the distance between them ($r$), the depth of the potential well
  ($\varepsilon$), and the fall-off distance ($\sigma$) of the interaction potential. Particle properties such as
  acceleration ($a$) or velocity ($v$) change according to the forces, 
  causing the particles to move. Additionally, a cutoff radius to ignore negligibly small
  long-range interactions is applied.
  
\begin{figure}[tp]
    \centering
    \begin{minipage}[b]{.51\textwidth}
      \includegraphics[scale=0.95]{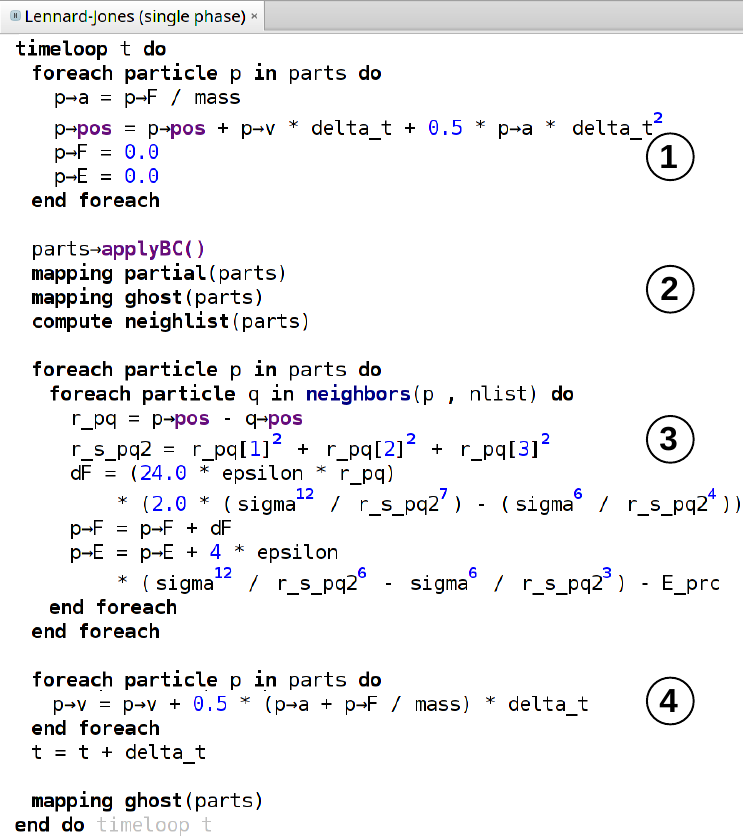}
      \caption{The simulation loop body for the Lennard-Jones dynamics.}
      \label{fig:ppme-lj-loop}
    \end{minipage}
    \hfill
    \begin{minipage}[b]{.48\textwidth}
      \centering
      \includegraphics{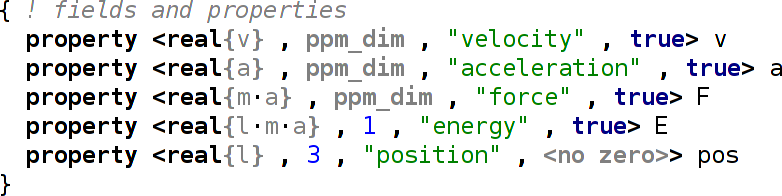}%
      \caption{Dimensions annotated to particle properties in the Lennard-Jones example.}
      \label{fig:dimension_annotation}
      \vspace{6mm}
      \includegraphics{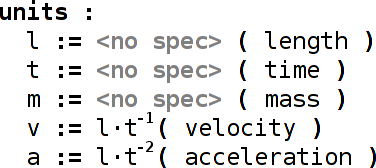}
      \caption{Declarations of base dimensions (length, mass, and time) with velocity and acceleration as derived dimensions.}
      \label{fig:dimension_declaration}
     \vspace{6mm}
     \includegraphics{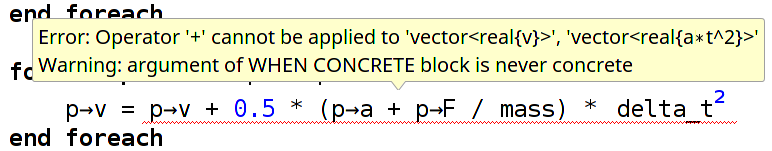}
      \caption{User notification of an error caused by incompatible
      dimensions.}
       \label{fig:ppme-type_dimension_error}
      \end{minipage}
  \end{figure}

    % JC: Next as separate paragraph - (issue 13) 
  The essential part of simulating the potential is located in the \inline{timeloop}
  depicted in Figure~\ref{fig:ppme-lj-loop}. Therein, the force acting on the
  particles due to pairwise interactions is computed and applied. The loop can be
  divided into four sections (\circled{1}--\circled{4}). First, the particle positions (\arrowOp{p}{pos}) are
  updated based on the values of velocity (\arrowOp{p}{v}) and acceleration
  (\arrowOp{p}{a}) (cf.~\circled{1}). After the particle positions change, the boundary
  condition must be imposed, followed by updating the mappings and neighbor list (cf.~\circled{2}). 
  The block of two nested particle loops implements the actual particle--particle 
  interactions (cf.~\circled{3}). For each particle $p$ the pairwise interaction with all nearby 
  particles $q$, retrieved via \inline{neighbors(p, nlist)}, is computed. The force
  $F=-\nabla E$ acting between two particles and the potential (or energy) $E$ are given by
  \begin{equation} \vec{F}(r) = 24 \varepsilon r \left( 2 \frac{\sigma^{12}}{r^7} -
  \frac{\sigma^6}{r^4}\right) , \qquad E(r) = 4 \varepsilon \left[ {\left(
  \frac{\sigma}{r}\right)}^{12} - {\left( \frac{\sigma}{r} \right)}^{6} \right]
  \end{equation}% 
  This corresponds to the lines with assignments to \inline{dF}
  and \arrowOp{p}{E}, respectively, where \inline{r_s_pq2} corresponds 
  to the squared distance $r^2$ between $p$ and $q$.
  The last update on the particle list modifies the
  velocity as a consequence of the new force (cf.~\circled{4}).
% paragraph lennard_jones_potential (end)

In this example, we also use dimensions to further improve static error detection.
A \emph{dimension declaration} in PPME resides in a special file owned 
by a model, where each declaration contains an identifier \inline{<d>}, an 
optional specification \inline{<spec>}, and an optional suggestive name \inline{<desc>}.
Figure~\ref{fig:dimension_declaration} shows the
declaration of fundamental and derived dimensions. From the three fundamental
dimensions length ($l$), time ($t$), and mass ($m$), convenient notations for
velocity ($v$) and acceleration ($a$) are derived. 
%IFS: maybe define a=v/t in order to show that dimensions can also be nested? 
Note that this specification
is not bound to this example and can be reused in any other PPME project.

% JC: Next as separate paragraph - (issue 13)
Figure~\ref{fig:dimension_annotation} shows the annotated particle properties,
which are used by PPME's type system to derive expression types. The  
type system corresponds to the formalization introduced in the previous section.
It enables capturing type and dimension errors right in the editor. Errors are reported 
to the user where they occur, as shown in Figure~\ref{fig:ppme-type_dimension_error}
using the deduction in Figure~\ref{fig:type_deduction}.
The outer addition is highlighted, and the information states that the operation cannot be applied
to operands with given (annotated) types  $[\real, v]$ and $[\real, a \cdot t^2]$. 
    
\subsubsection{N-body Simulation} % (fold)
\label{par:n_body_simulation}
As a third case study, we implement an N-body simulation of two galaxies using 
particle methods. As the model structure of this example is similar to the 
Lennard-Jones example above, we skip the corresponding details
of the code and focus on another important aspect of PPME: its interface with the
underlying Fortran language. 

% JC: Next as separate paragraph - (issue 13) 
\revii{PPME is designed as a standalone DSL, nevertheless unanticipated 
use cases can be supported by inline code statements. In the N-body simulation, 
the initial particle data need to be loaded from an external data source \inline{data.tab}.
As PPME has no built-in functionality to import data from this type of file, it has to 
be specified using custom code.}
\begin{figure}
 \centering
    \includegraphics{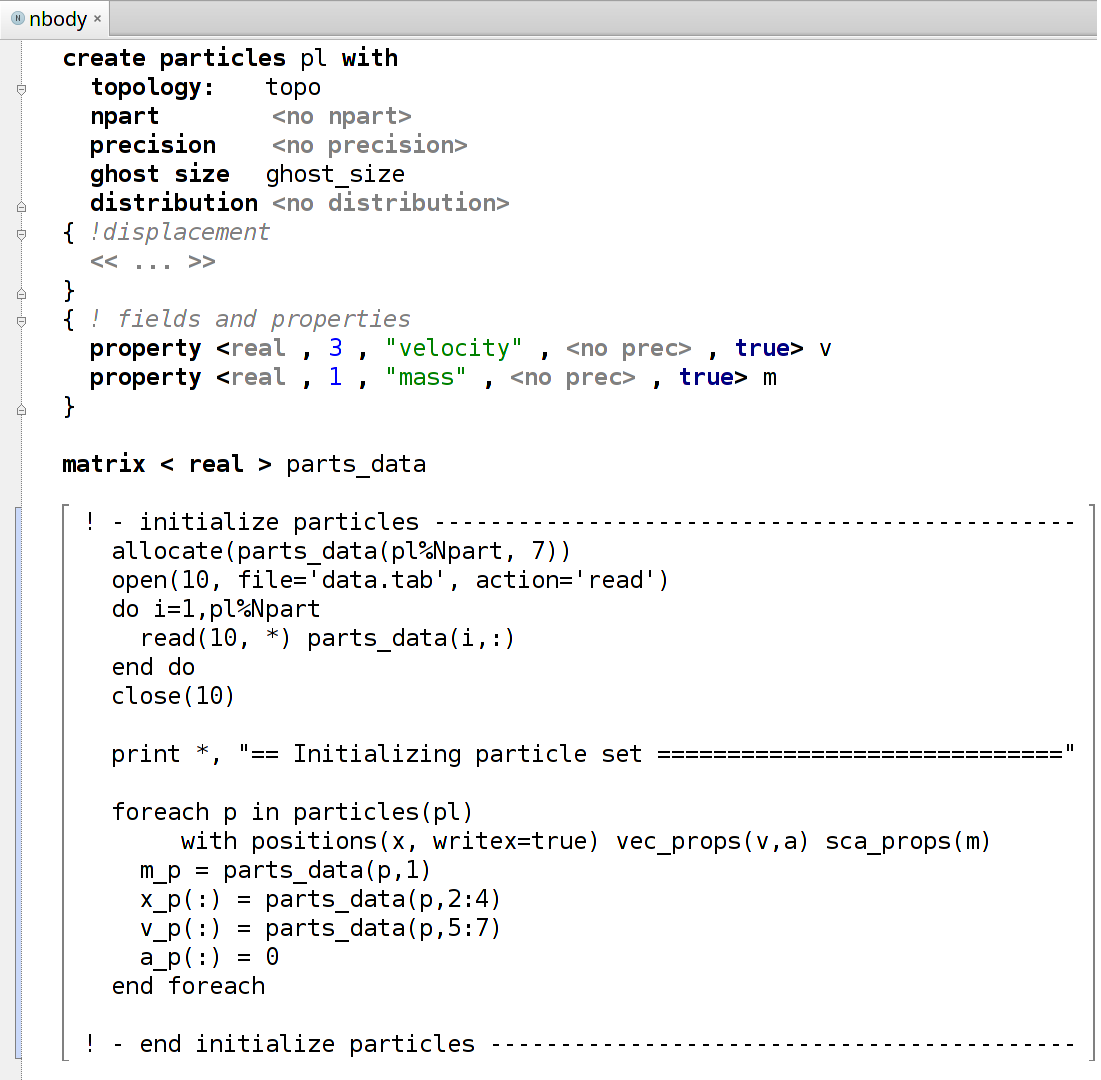}
    \caption{Particle initialization using PPML inline code.}
    \label{fig:ppme-nb}
\end{figure}
This can be achieved through an \inline{InlineCodeStatement}, which supports 
inlining arbitrary Fortran or PPML code directly into the program. Figure~\ref{fig:ppme-nb} shows how
\inline{data.tab} is read and into PPME's data structures. The PPML code is located 
within a pair of squared brackets, which demarcate it from the rest of the program. 
The code has direct access to the matrix \inline{parts_data} as well as to the declared 
fields \inline{v} and \inline{m}. During this custom initialization, the data are first loaded into 
the matrix (which corresponds to a Fortran array) and afterwards copied to the corresponding
fields. 

Notice that the inlined PPML code is not analyzed by PPME. Errors may therefore be 
introduced by the developers that propagate to later stages in the compile chain. 
However, such code can be conceptualized easily by extending the language and
converting it into an MPS generator, which is one of the central ideas in 
language-oriented programming.

% paragraph n_body_simulation (end)
% subsection case_studies (end)

% section the_ppm_environment (end)

% vim: set ts=2 sw=2 sts=2:
% vim: set wrap breakindent:
% !TEX root = paper_ppme_TOMS.tex

\section{Numerical Optimizations and Tool Integration}
\label{sec:analysis}
Applications in science and engineering often depend on floating-point
arithmetic in calculations to approximate real arithmetic. In this section, we
therefore introduce an \revii{accuracy} optimization for floating-point expressions that is
integrated into PPME. 
This serves as an example of how to extend PPME by existing tools, 
\revii{thereby demonstrating the possibilities offered by the high-level 
expressions of the language}. 
In a similar way, other common optimizations can be added in the future, 
for example loop transformations to improve data locality and 
performance~\cite{Lam1991,luporini_coffee_2015}.
\revii{Loop optimizations are orthogonal to the floating-point 
transformation described in this section}.

%As numerical operations are such an essential part of scientific computations they
%are a common target for analysis and optimization. Likewise, focus can be put on
%numerical optimizations based on transformations of floating-point expressions.
In the case of floating-point computations, a compiler can trade \emph{accuracy}
for \emph{performance}. Such optimizations often rely on
\emph{abstract interpretation} for preserving semantics. The abstract semantic can be
used to build program equivalent graphs, inspect them with regard to the desired
optimization target, and enable efficient detection of appropriate
rewrites~\cite{Ioualalen2012}. 
As an example of such optimizations, we adopt
\emph{Herbie}\footnote{\url{https://github.com/uwplse/herbie}} as a recent approach to 
automatically improve floating point expressions~\cite{Panchekha2015}. 
Herbie relies on heuristics to estimate and localize rounding errors at sample points. 
%
%To improve a given expression, it applies a
%database of rules, takes series expansions, and combines improvements for different
%input regions. After initialization, the tool localizes floating-point inaccuracies
%through sample point computations. 
%
Once low accuracy (e.g., numerical extinction) is detected, Herbie attempts to improve the program 
by rewriting inaccurate expressions using a rule database. 
%
%The error elimination is carried
%out by applying rewrite rules describing basic arithmetic artifacts. 
%
Thereafter, if possible without loss of accuracy, the expressions are simplified. 
Finally, Herbie may apply series expansions for inputs around zero or near infinity to better 
approximate the result.
%
%In case rounding errors are detected for
%inputs around zero or near infinity for which no better program could be found using
%rewrite rules, Herbie performs series expansions to approximate the result. This
%often helps to avoid over- and under-flows. 
%
This process is repeated iteratively so that new candidate expressions are yielded after
each iteration, keeping only the programs which achieve the best accuracy at least
at one sample point. Finally, Herbie uses one or more candidates to achieve an improvement
of accuracy over all sample points.
%As in most cases not a single candidate program is most
%accurate for all input points, Herbie assembles a final program in a regime-inference
%pass. Regimes correspond to input regions for which different expressions are most
%accurate. In order to prevent over-fitting, regime inference has to find a trade-off
%for the number of branches to apply. Hence, Herbie combines several candidate
%programs to achieve an improvement of accuracy over all input points.

\subsection{Tool Integration}
We integrated Herbie as a \revii{plugin into PPME. MPS provides convenient configuration languages to define \emph{plugin solutions} and to specify their behavior. The solution of our Herbie plugin comprises a preference page, an action object that executes Herbie in a separate process, and a \emph{mapping script} that collects references to expression nodes in the PPME program.}
Since the algorithms \revii{for accuracy optimization} are computationally expensive, expressions are not 
inspected automatically but must be flagged for evaluation by the user. Users can 
trigger the analysis and transformation process for the active PPME editor and the 
result is then annotated to the corresponding fragments in the code.

\begin{figure}
\begin{minipage}{.42\textwidth}
  \includegraphics[width=\textwidth]{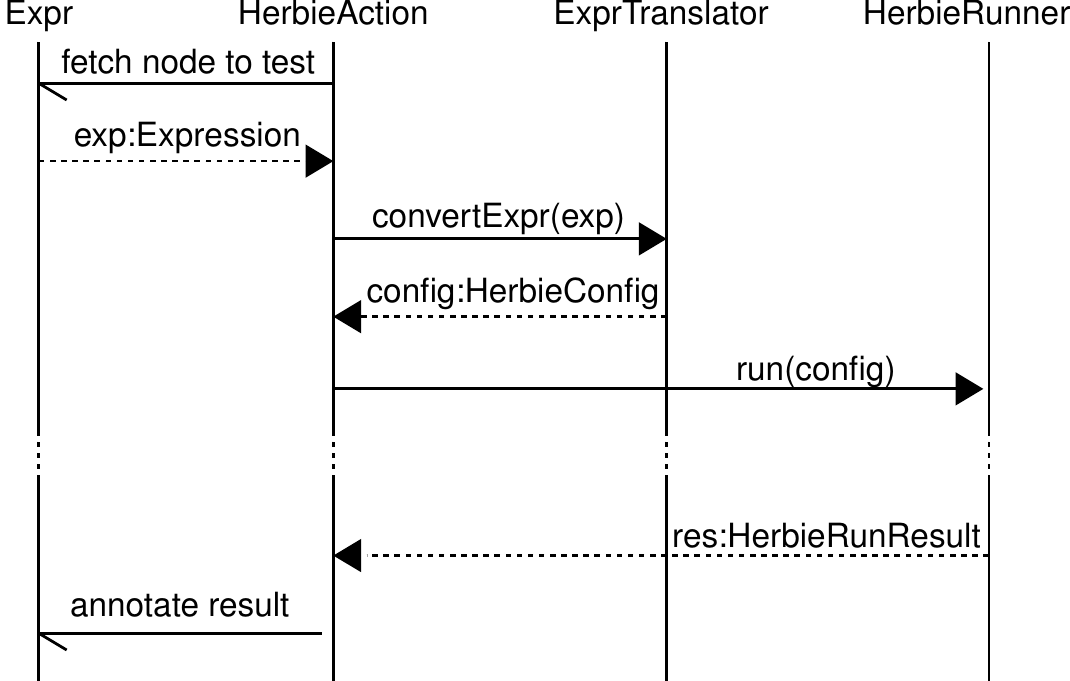}%
  \caption{Sequence diagram for the execution of a Herbie analysis for a single expression.}%
  \label{fig:ppme-herbie-seq}%
\end{minipage}
\hfill%
\begin{minipage}{.55\textwidth}
  \includegraphics{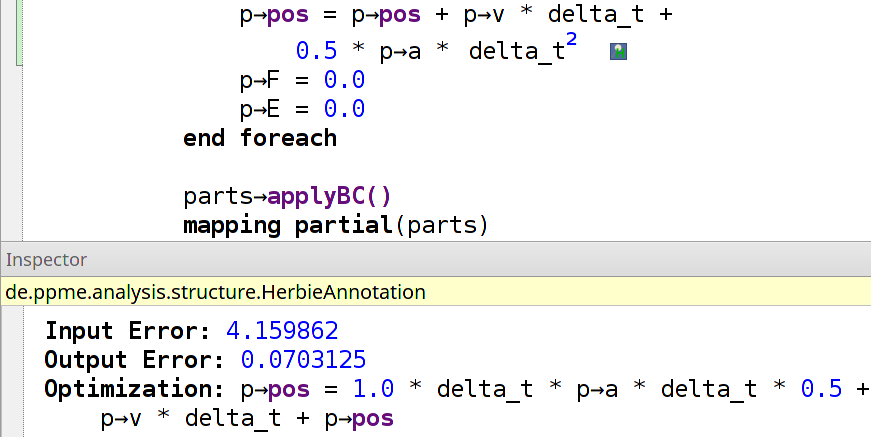}%
   \caption{Inspection view of an annotated expression with the results reported by Herbie.}%
   \label{fig:ppme-herbie-intention}%
\end{minipage}
\end{figure}

%Herbie requires its input to be in the form of a \inline{herbie-test} instance in
%the Racket language:
%%  %
%  \begin{lstlisting}[basicstyle=\small\ttfamily, numbers=none, framesep=0pt,
%  rulesep=0pt, frame=none, mathescape=true] (herbie-test (<variables>) <name>
%  <expression>) \end{lstlisting}
% %  %
%  Each test has an arbitrary name field (\inline{<name>}). Commonly, a combination
%  of filename and line number of the expression to test is used. Since simulations
%  written in PPME are no traditional text files using the node id of the root
%  expression to test is a sane choice. As a benefit, this allows to map a given test
%  case back to a specific node in the PPME program. Besides the name tag, all free
%  variables in the expression \inline{<expression>} are listed in the variables
%  block \inline{<variables>}. The free variables are the variables to test in the
%  Herbie execution, i.\,e., these variables are tested with a series of sample
%  points.
% 
Expression annotation works via the MPS intentions dialog (Alt + Enter). 
%The intention will always annotate the root of the expression the cursor is currently placed on. 
A small icon indicates that an expression is marked for analysis (cf. first line in 
Figure~\ref{fig:ppme-herbie-intention}). 
All marked expressions are transformed into a prefix notation that matches Herbie's 
input language \emph{Racket}, built-in operations and data types.
PPME maintains the translated expression as a string, a reference to 
the original expression node and its identifier, as well as a table of variables. 
For the expression highlighted in Figure~\ref{fig:ppme-herbie-intention} the following test case is
generated:
\begin{lstlisting}[basicstyle=\footnotesize\ttfamily, language=lisp,morekeywords={herbie-test},numbers=none, framesep=0pt, rulesep=0pt, frame=none, mathescape=true]
(herbie-test (p_pos p_a delta_t p_v) "2430378650379961582"
  (+ (+ p_pos (* p_v delta_t)) (* (* 0.5 p_a) (expt delta_t 2))))
\end{lstlisting}
where the \texttt{herbie-test} macro is called with a list of variables, a unique name, and
the actual translated expression.

Figure~\ref{fig:ppme-herbie-seq} illustrates the whole process (steered by 
\inline{HerbieAction}) responsible for fetching the nodes to test, analyze the node, 
and write the result back as a UML sequence diagram.
The result of a Herbie execution is summarized and displayed to the user in the
inspection view as shown in Figure~\ref{fig:ppme-herbie-intention}. The original
expression is annotated with additional information, i.e., the error of the input and output
expressions, and the optimized computation regime for the arithmetic expression. This
allows inspecting the result generated by Herbie, and it keeps the
original expression in place without modifying it. Instead, PPME users
are responsible for replacing the annotated node by the optimized one before the
text generation phase. This prevents unwanted expression rewriting. 
Note that the generic design of the
execution model and the result container allow reusing the same setup for other
tools by extending the existing framework.
%Instead of transforming the original
%node into target source code the annotated optimization has to be generated. 
Moreover, analysis and code generation remain separate
processes. 

\subsection{Accuracy \revii{Optimization}}
\label{par:accuracy_improvements}

\begin{figure}
\footnotesize
\improvementsFigure%
{7.1231523}%
{0.08203125}%
{%
\[
\frac{\partial c \to U}{\partial t} = D_u \cdot \nabla^2 c \to U - c \to U \cdot c \to V^2 + F \cdot (1.0 - c \to U)
\]
\vspace*{-8pt}
\lstinputlisting[language=diff, basicstyle=\scriptsize, linerange=135-137, numbers=none, nolol,frame=tblr]  % chkTeX 8
{img/diff_gs_orig_herbie_2.diff}%
}
\caption{Exemplary improvements for an expression taken from the Gray-Scott example.}
\label{fig:herbie-improvements:gs_dU} % chkTeX 24
\end{figure}
We investigate the improvements in accuracy for two of the case-studies presented 
in Section~\ref{sub:case_studies}, the Lennard-Jones
simulation (LJ) and the Gray-Scott reaction-diffusion system (GS). We annotate
several expressions in each program, and execute them with and without optimization.
Figure~\ref{fig:herbie-improvements:gs_dU} shows the analysis result for 
$\frac{\partial U}{\partial t}$ for the GS example, including
input and output error, the original expression, and the difference in the
generated source code. The input expression has an average error of
 seven bits (cf.~\cite{Panchekha2015}), 
and Herbie was able to nearly remove inaccuracies by 
expanding and redistributing terms. We compare the numerical results for
simulations with $t_{start}=0$, $t_{end} = 4000$ and $\Delta{t}=0.5$. 
The computed values for $U$ and $V$ differ in the last four to seven of seventeen significant digits, 
which confirms that the changes have an impact. In our visualization (cf. Figure~\ref{fig:grayscott-ppml-sim}), 
the differences are not noticeable. However, a longer simulation time may yield visible differences.

For the LJ case study, we investigate several expressions, this time with and 
without considering known restrictions on input values. \revii{For example,} consider 
Equation~\ref{eq:lj-03}.
\begin{align}\label{eq:lj-03}
	dF &= (24.0 \cdot \varepsilon \cdot r_{pq}) \left(2.0 \cdot \frac{\sigma^{12}}{r_{s_{pq}}^{7}} - \frac{\sigma^{6}}{r_{s_{pq}}^{4}}\right)
 \text{, \footnotesize where $r_{s_{pq}} > 0, \sigma \in [10^{-2}, 10^{-1}], \varepsilon \in [10^{-14}, 10^{-13}]$} 
\end{align}
In the case without value-range restrictions, the analysis found an improvement of 
${34.0}\mapsto{15.6}$ bits. However, this theoretical potential is
not reasonable when considering actual variable ranges, since the algorithm checks 
the whole domain of input values instead of optimizing over a small feasible interval only.
Consequently, only one of the analyzed expressions yielded 
an actual improvement after accounting for additional constraints using range annotations
(e.g., parameters with constant values) obtained from PPME's code analysis. 
As a consequence, the analysis did not find significant
improvements\footnote{In fact, the analysis increased the error from $5\mapsto{11}$ ($5\mapsto{20}$) bits 
on Racket version $6.4$ ($6.7$) using seeds $2808995595$, $415209655$, $1218262282$, $3135925998$, $2713258581$,
$1066853958$ and Herbie commit hash \texttt{f6ebaea}. In contrast, in version $1.0$ of the tool, input and output error
remained at around $4$ bits.}. 
Hence, additional information about variables may be required to generate reliable results. A DSL like ours
may help extract such information automatically.  

\subsection{\revii{Impact} on Runtime Performance}
\label{par:influence_on_runtime_performance}
Since the optimization modifies expressions and, in some cases, replaces a simple assignment
with a complex one containing several conditional branches, its influence on runtime
performance might be of concern. Therefore, we investigate the impact on execution time 
for the GS and LJ case studies. We compare the runtime of
the original program for each use case with the optimized versions. To factor out 
disk-I/O from the measurements, the
simulations are modified so that no output is generated. The tests
were run on a system with an Intel Core i3-4160 CPU, 16~GB random-access memory 
and Ubuntu Linux with kernel 4.2.0.
%\footnote{While this is not an HPC machine, we found this 
%setup sufficient for our experiments since we were not interested in showing absolute 
%performance numbers but what kind of impact can be expected from such kinds of 
%accuracy optimizations.}

\begin{figure}[tp]
	\begin{minipage}[b]{.45\textwidth}
		\centering
		\includegraphics[width=1\textwidth]{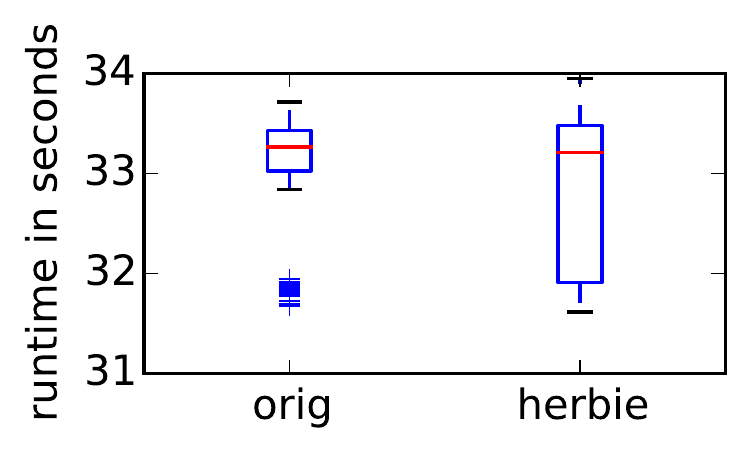}
		\caption%
		[Runtime comparison for the Gray-Scott case study.]
		{
		Runtime comparison for the Gray-Scott example with $t_\mathrm{end} = 2000$.
		The median runtime for both simulations is nearly identical, which indicates that Herbie's optimization have no impact on the program's runtime performance.
		}
		\label{fig:bench_gs}
	\end{minipage}
	\hfill
	\begin{minipage}[b]{.45\textwidth}
		\centering
                 \includegraphics[width=1\textwidth]{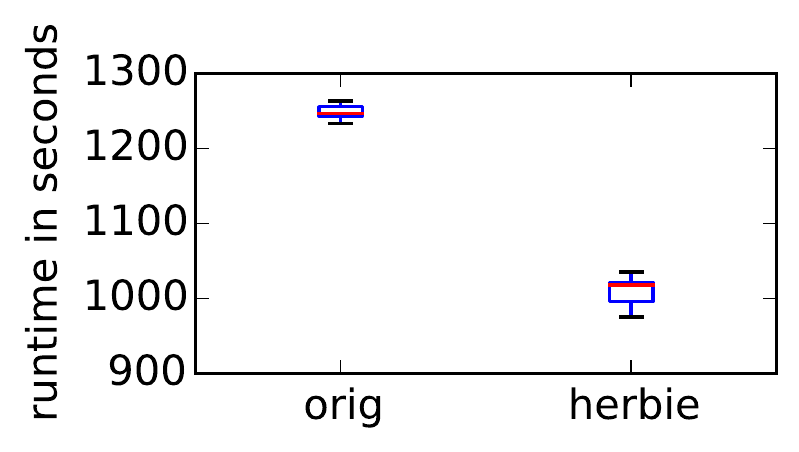}
		\caption%
		[Runtime comparison for the Lennard-Jones case study.]
		{
		Runtime comparison for the Lennard-Jones example.
		The execution of the program modified by Herbie was approximately $20\,\%$ faster than the original implementation.\newline
		}
		\label{fig:bench_lj}
	\end{minipage}
\end{figure}

We executed the GS use case $100$ times per variant with $4000$ steps 
($t_{start} = 0.0$, $t_{end} = 2000.0$, $\Delta{t} = 0.5$).  Figure~\ref{fig:bench_gs} 
shows the variation of the execution times as a box plot. The median of both 
variants is nearly identical while the data are less scattered for the original simulation
with a few outliers at approximately the minimal execution time of the modified 
program. For GS there is therefore no significant performance impact due to the accuracy 
optimizations.
%\sk{Note: to further verify this, we could rerun the experiments 1000 or 10000 
%times, this could also help to get rid of the outliners}

In the case of LJ, we compared the runtime of the original program 
to a variant with several transformed expressions, including Equation~\ref{eq:lj-03} 
without range restrictions. The simulation is  
executed for $n = 5000$ particles, end time $t_\mathrm{end} = 0.2$, and 
$\Delta{t} = 1.0 \cdot 10^{-6}$ ($200,000$ steps). The results are summarized in
Figure~\ref{fig:bench_lj}, both variants were run $25$ times.
%IFS: what was sigma and epsilon?
%
The accuracy-optimized version runs nearly $20\,\%$ faster than the
original implementation. However, this can be attributed to over-simplifications of
some of the expressions due to the missing range restrictions. Considering the actual numbers
produced by the simulation, the two variants visibly differ in their results, with the optimized version being less accurate, \revii{yielding force values that are two orders of magnitude lower. This is the result of Herbie removing a complete subterm from the force equation.}
When taking parameter value ranges into account, only one expression could be improved 
through a simple restructuring of its terms. In this case, we can not
detect any impact on the program's execution time, but the results remain correct.

% vim: set ts=2 sw=2 sts=2:
% vim: set wrap:

\section{Evaluation}
\label{sec:evaluation}
%In this section we evaluate PPME w.r.t. the major challenges it was meant to address.
%
%\paragraph{Reducing the Knowledge Gap}
\label{par:evaluation-reducing_the_knowledge_gap}
One of PPME's primary goals is to reduce the knowledge gap in scientific programming.
This is achieved by providing domain-specific abstractions at the language level for
particle-mesh simulations, based on those previously offered by PPML. This is
complemented with features of a modern IDE, such as code completion and syntax highlighting,
guiding the scientific programmer using domain-specific notations 
that are free of the overhead otherwise introduced by parallel programming.  
In addition, the formal type system and its optional
extensions considerably improve error detection. They prevent a series of common
errors at development-time and provide developers with meaningful feedback. Errors
are captured and reported at the DSL level, instead of the level of the generated code.
In comparison with PPML, PPME improves error detection and handling of the following kinds:
\begin{itemize}
\item \emph{PPM instantiation errors.} PPME inherently generates statements in the order
that is expected by the PPM call protocols. Frequently, such errors would otherwise only be 
discovered at runtime.
\item \emph{PPML redundant redeclaration errors.} PPML requires a proper redeclaration of
fields and operators that are accessed in loops or right-hand-side specifications. A missing 
or wrong declaration leads to compile-time or runtime errors. Since PPME
analyzes the code to derive the required information, such errors become impossible.
\item \emph{Syntax errors in Fortran code.} Except for explicitly inlined Fortran code, PPME 
has its own expression language so that syntactic errors in Fortran expressions, as in PPML,
are impossible.
\item \emph{Type errors in Fortran code.} PPME has its own type checker so that static 
type errors in the generated Fortran code are not possible.
\item \emph{Dimension-related errors.} Due to dimension support, PPME is capable of 
statically detecting errors in expressions and differential equations. If not detected, such errors
can silently corrupt the simulation result, wasting HPC resources.
\end{itemize}

  %
  % Counting WLOC (Written Lines of Code) for better metric on the implementation effort:
  %
  % ------+-------+-------+--------
  % WLOC  | PPME  | PPML  | Fortran
  % ------+-------+-------+--------
  % GS    | 28    | 53    | 623
  % LJ    | 72    | 75    | 480
  % NB    | 51    | 58    | 469
  % ------+-------+-------+--------
  %
  \begin{figure}[tp]
    \centering
    \footnotesize
    \begin{tabular}{l c c c c}
      \toprule
        & PPME  & PPME*   & PPML  & Fortran   \\
      \midrule
      Gray-Scott (complete)
        & 46    & 28      & 53    & 623       \\
      Gray-Scott (RHS)
        & 4     &  4      & 9     & 103       \\
      \midrule
      Lennard-Jones (complete)
        & 88    & 72      & 75    & 480       \\
      \midrule
      N-Body (complete)
        & 69    & 51      & 58    & 469       \\
      \bottomrule
      \multicolumn{5}{r}{\scriptsize PPME*: Written Lines of Code (WLOC)}
    \end{tabular}
    \caption{Comparison of lines of code for PPME, PPML, and PPM/Fortran.}
    \label{fig:comparison-loc}
  \end{figure}
Besides error detection, PPME achieves program sizes similar to PPML or even smaller. Figure~\ref{fig:comparison-loc} compares the source lines of code (SLOC)
of the implemented case studies for PPME and the generated PPML and Fortran codes.
Since PPME is not a conventional editor but a projectional one, we use 
\emph{written lines of code (WLOC)} as a second metric in column \emph{PPME*}. 
This takes into account that PPME programs typically contain some lines that have not been 
entered by the developer, but are generated by the editor (e.g., optional configuration 
fields). 
Considering SLOC, in the Lennard-Jones and N-body examples, the PPME programs are
larger than the generated PPML code with ratios of $88:75$ ($117\,\%$) and $69:58$ ($119\,\%$).
In contrast, in the Gray-Scott example, PPME requires less space with a ratio of $50:62$ ($81\,\%$).
The reduction in the latter case is due to the built-in constructs for solving PDEs, which
are not used in the other examples.
Considering WLOC, PPME reduces the code sizes in all examples. In the 
Lennard-Jones and N-body examples, the corresponding ratios versus PPML go down to $96\,\%$ and 
$88\,\%$, respectively. For Gray-Scott, the code size goes down to $52\,\%$.

\revii{In terms of performance, there is no difference between the execution time 
of code generated from PPME and an equivalent hand-written PPML version. 
This is due to the fact that the PPML output code generated from PPME is identical to the hand-written PPML code.
Any performance difference between PPML and plain Fortran linked against the PPM Library was found to be
less that $2\,\%$ in previous experiments~\cite{Awile:2010a}, whereas PPM Fortran was 
found to generally perform better than hand-parallelized Fortran code that does not use the PPM Library~\cite{Sbalzarini2006a}.
} 

\revii{PPME also has some drawbacks. In comparison to PPML and other programming languages, 
version control turns out to be more complicated. Since files are serialized using XML, 
conventional text-based diff and merge operations are difficult to apply. While
MPS has built-in support for most of the established version-control systems, 
the resulting workflow is different from the text-based approaches and not always as 
efficient. For instance, even if the rendered program  did not visibly change, or only a small 
edit was applied,  serialization may change a lot, causing more merge conflicts in 
collaborative development scenarios.}

\revii{Another difference between PPME and general-purpose languages is that convenience support for user-defined 
functions and types is not built into the PPML language. For the development of our current application examples,  
it was not necessary to have these concepts in the language. However, this may change in the future, demanded by 
more complex applications. Language extension is one of the key features of language-oriented programming 
in PPME/MPS, which was, in turn, not the case for PPML.}

\section{Related Work}
\label{sec:related_work}
%language workbenches and viewbased editing

%Language workbenches aim at providing integrated and holistic language development
%environments for DSLs. They achieve this by combining parsing, rewriting, and
%analyses approaches with automatically generated editing services for DSL-specific
%IDEs, e.g., syntax highlighting, code completion, refactoring and debugging support
%Hence, the effort for implementing and testing of DSLs is drastically reduced. 
An exhaustive
overview of language workbenches, their features and use is given
in~\cite{erdweg_workbenches_2013}. Here, we therefore only discuss a small selection
of well-known workbenches for textual languages. 

Spoofax~\cite{kats_spoofax_2010} is a language workbench that builds upon term
rewriting with {\it Stratego}~\cite{bravenboer_stratego_2008}, a high-level grammar
language as well as meta languages for name and type analysis. DSLs implemented in
Spoofax can be used via generated plugins for the Eclipse platform or from command
line. Other well-known workbenches for textual DSLs with similar capabilities on the
basis of the Eclipse Modeling Framework (EMF) are Xtext~\cite{eysoldt_xtext_2010} and
EMFText~\cite{heidenreich_derivation_2009}. These tools leverage the relation of
context-free grammars and the EMF (cf.~\cite{alanen_ebnf_mof_2003}) to make
grammarware available to the field of model-driven software engineering. Xtext
provides built-in languages for code generation and semantic functions. Furthermore,
an Xtext language for formal specifications of type systems has been
developed~\cite{bettini_implementing_2015}. EMFText, in contrast, follows a
\emph{convention-over-configuration} approach, which tries to provide most language
features out of the box. In case that this is not sufficient to realize all intended
behavior, model-based attribute grammars or the object-constraint
languages~\cite{buerger_reference_2011,heidenreich_model-based_2013} are available.

\revii{Since MPS implements a projectional editing approach~\cite{feiler_incremental_1981,voelter_mps_2013}, no parsers or grammars are involved, as nodes are added, deleted, and modified directly.} While projectional editing is more restrictive
than direct text manipulation, it is less prone to errors and serializes models as
data structures, i.e., when a model is saved and loaded again, the exact instance is
restored. 
%Furthermore, it allows for advanced rendering of tables and mathematical
%equations, which is an advantage in scientific computing. 
%Like the other language workbenches, MPS provides a collection of built-in languages
%to cover typical DSL development aspects, such as code generation and type analysis.
MPS has already proven its applicability in other domains. The \emph{mbeddr} project instantiates
MPS in the embedded domain providing a projectional C frontend and several extensions
and domain-specific analyses, such as state machines and model checking~\cite{Voelter2013}. 
Moreover, the authors of \cite{Benson2015} used MPS to create a language and
editor for automated statistic analyses of biological data (bio markers), which is 
designed for end-user programming and statistical visualization. 

%extensible compilers/languages
Besides language workbenches and greenfield DSL development, another possibility is
to hook into already existing extensible compiler infrastructures, e.g., relying on
their basic intermediate representations, default optimizations, and code generation
facilities. Well-known examples for these infrastructures are the LLVM
framework~\cite{lattner_llvm_2004} and Graal~\cite{duboscq_graal_2013}. 
LLVM is used as a compiler backend for various general-purpose languages, most notably C and C++. It is
centered around a universal intermediate language that is transformed through several
extensible phases, as well as various optimizations, allocations, and code selection,
down to platform-specific machine code. A DSL could rely on this infrastructure by
generating code in the LLVM intermediate language, reusing and extending compiler
facilities.
Graal is an extensible just-in-time compiler for the Java Virtual Machine and a
platform for testing new high-level optimizations. Further, it provides support for
integrating with new languages, language features, and domain-specific
optimizations~\cite{wimmer_graal_2015}.

% DSLs in scientific computing
During the last years, the importance of DSLs for scientific computing has been increasingly 
realized. This led to the emergence of a number of approaches of which
we mention a few notable examples. Blitz++~\cite{veldhuizen_blitz_2000} is a
template-based library and DSL for generating finite-difference operators (stencils)
from high-level mathematical specifications. Freefem++~\cite{hecht_freefem_2012} is a
software toolset and DSL for finite-element methods. This DSL allows
users to define analytic as well as finite-element functions using domain
abstractions such as meshes and differential operators.
Liszt~\cite{devito_liszt_2011} extends Scala with domain-specific statements
for defining solvers for partial differential equations on unstructured meshes with support for parallelism through MPI, pthreads, and CUDA. The
FEniCS project~\cite{logg_automated_2012} comprises a finite element library, the
unified form language (UFL)~\cite{alnaes_ufl_2014}, and several optimizing compilers
for generating code that can be used with the library. Building upon FEniCS, the
Firedrake project~\cite{rathgeber_firedrake_2015} adds composing abstractions such as
parallel loop operations.
%Similarly, in the future, we plan to develop a set of particle-method related optimizations 
%on top of PPME. For mesh-based discretizations, it would be an opportunity to integrate 
%the UFL, providing a unified framework for both abstractions.
%IFS: I would not mention this. I also think UFL might not be good because it is centered on bilinear forms and weak discretizations, which you only get in FEM

% DSL optimizations
%In many cases, it suffices to translate the DSL code into a lower-level target language like
%C and let its compiler apply standard optimization. 
%Otherwise, rules for optimizations may be programmed or realized using some of the approaches discussed
%previously.
The idea of transforming or rewriting program code for optimization purposes is not new. 
For example, a DSL optimizer could be implemented using program
transformations~\cite{schordan_user_optimizations_2003} or rewrite rules. However,
research on using graph-rewrite systems for such
tasks~\cite{amann_graph_2000,schoesser_graph_optimizations_2008} indicates that the
pattern language must be powerful enough to express context-sensitive patterns.
Furthermore, recent research shows that rewriting is a convenient technique for
implementing high-level optimizations on a restricted set of language constructs. For
instance, authors in~\cite{Panchekha2015} propose a method for
automatically improving the accuracy of floating-point expressions by rewriting such
expressions according to a set of harvested patterns. Further, the authors
of~\cite{steuwer_rewriting_2015} apply rewriting to specific functional expressions
for parallel computations to obtain efficient GPU kernels.
In the field of DSLs for scientific computing, domain-specific optimizations carry
great potential since scientific codes often induce specific boundaries on data
access and numeric algorithms. In~\cite{olgaard_fem_optimization_2010}, the authors
discuss different optimization strategies on representation code for element tensors
in the finite-element method. The representation code is written in
UFL, a high-level mathematical DSL for variational forms. The proposed strategies yield significant runtime speedups and leverage
domain knowledge to automate nontrivial optimizations that normally would have been
developed manually by scientific programmers. Related to that, the authors
of~\cite{luporini_algorithm_2016} discuss loop-level optimizations for finite-element
solvers in the COmpiler For Fast Expression Evaluation
(COFFEE)~\cite{luporini_coffee_2015}. Heuristics are used to predict local minima
of operation counts at runtime, using semantics-preserving transformations such as
code motion, expansion, and factorizations. The authors show that their
domain-specific optimizations are superior to those that are generally applied by
standard compilers such as Intel's ICC for optimizing the operation count in nested
loops. Similar optimizations could be provided as extensions of PPME.
%\jc{Here again a statement about us is needed. Something like, this could be integrated 
%in our super flexible environment ;-)}

% Units/dimensions
Also, the idea of adding dimensions or physical units to DSLs is not new.
\cite{Cook2006} presented an analysis technique checking correctness of 
units in programs without extending the base language, aiming for a minimal effort of 
annotations for a developer. Furthermore, \cite{Austin2006} proposed unit annotations 
for linear-algebra and finite-element calculations, which are similar to the dimension annotations
in PPME.
However, adding units to programming languages frequently has flaws. 
For instance, frameworks may use abstractions with boxing and unboxing of 
quantities and units, which implies a runtime overhead.
In our approach, the analysis is optional and does not have an impact on runtime 
performance, since it is only used at compile-time for consistency checking and does not persist in the simulation. 

% SK: this sentence is strange, units/dimension do not persist in the simulation, right?
%The core language is
%independent of the extension, meaning that a developer can chose whether to use
%dimensions in a simulation or not, and that existing programs can be annotated with
%unit information subsequently.

\section{Conclusions}
\label{sec:conclusions}
We have presented PPME, an adaptable and extensible programming environment environment with a domain-specific language for particle-mesh methods.
It aims to simplify the development of scientific simulations through domain-specific abstractions and automatic generation of client code that links with the PPM library.
Leveraging the language workbench MPS, we earned features that are typical for 
modern development environments (e.g., syntax highlighting, automatic code completion, etc.) and
also features that are due to the methodology of projectional editing (e.g., mathematical 
notation). Furthermore, MPS provided us with a type-system language and a powerful
concept of arranging our environment in a modular way, which also is one of the key enablers 
for extensible language design and implementation, one of the major goals of PPME.

We demonstrated PPME's capabilities in that respect
in two ways: First, we developed a dimensional calculus on top of the original type system,
including an extension for checking and declaring dimensions, or even measurement units.
Errors discovered by the type system and the dimensional analysis are instantaneously reported to the user at design time. 
Second, we added support for automatic accuracy improvements of floating-point
expressions by adopting Herbie as an external tool and integrating additional 
value-range specifications into the language. Since both extensions are designed  
as independent plug-in solutions, they do not interfere with the base language and
the framework can easily be adapted to other cases, and developers are free to use 
the extensions only if desired.

Despite the obvious advantages of PPME, there are some obstacles that derive from MPS' basic principles of projectional
editing and modular language specification. 
Due to the complexity of MPS, it is not easy for scientific programmers to 
%MPS has a steep learning curve. The amount
%different language aspects and specification languages easily overwhelms new developers, 
%frequently similar concepts behave differently in different languages. Hence, if scientific 
%programmers intend to 
develop own extensions for PPME (e.g., for loading data from
a specific type of file). They have to become familiar with the concepts of MPS, which re-iterates 
the problem of the knowledge gap in scientific programming. PPME therefore allows the user to include custom Fortran code as inline blocks. While strictly speaking this is a design breach in a non-embedded DSL, it offers a pragmatic solution. Other issues with MPS come from 
projectional editing itself, which leaves less freedom than text editing w.r.t.~writing
comments or incomplete intermediate code. However, after some training, developers normally
get used to the tool (cf.~\cite{Voelter2013}). Another potential source of problems is that
programs are not stored as plain text, so that using version control outside of PPME/MPS is difficult.

In the future, we will extend PPME to support more 
particle and mesh abstractions, including inter-particle {\em connections}, {\em neighbor lists}, and {\em meshes} of different topology. We also want to expand PPME to better support 
high-level parallelization constructs and analyses to further improve code generation and 
runtime scalability by leveraging the domain knowledge for more intelligent mapping 
and distribution of computations onto underlying parallel hardware.
Finally, we will improve PPME's code generation process by adding another layer of abstraction
to better integrate the target language and make the backend exchangeable. This way, we 
will be able to support the successor of the PPM library, OpenFPM, which is currently developed in C++.

%\section*{Acknowledgements}
%This work is partly supported by the German Research Foundation (DFG) within the Cluster of Excellence “Center for Advancing Electronics Dresden”.

% --------------------------------------------------------------------------- %
% Bibliography
% --------------------------------------------------------------------------- %
\bibliographystyle{ACM-Reference-Format-Journals}
\bibliography{newbib,online}

\end{document}